%% file: paper.tex
\documentclass[aps,pre,showpacs,showkeys,twocolumn,superscriptaddress,
floats,floatfix,preprintnumbers,longbibliography]{revtex4-1}
\usepackage{amssymb,amsmath,amsthm}
\usepackage{bm,url,graphics,subfigure,epsfig,rotating,float}
\usepackage{soul}
\usepackage{mathrsfs}
\usepackage{hyperref}
\usepackage[dvipsnames, table, xcdraw]{xcolor}
\usepackage{color,soul}
\usepackage{relsize}
\usepackage{colortbl}
\usepackage{multirow}
\usepackage{bigstrut}
\usepackage{slashbox}
\usepackage[normalem]{ulem}   
\graphicspath{{./},{./pics/} }
\input{symdef.inc}
\begin{document}
\title{Flexible filament in time--periodic viscous flow :
  shape chaos and period three}
\author{Vipin Agrawal}
\affiliation{Nordita, KTH Royal Institute of Technology and
Stockholm University, 12 Hannes Alfv\'ens v\"ag
10691 Stockholm, Sweden}
\affiliation{ Department of Physics, Stockholm university, Stockholm.}
\author{Dhrubaditya Mitra}
\email{dhruba.mitra@gmail.com}
\affiliation{Nordita, KTH Royal Institute of Technology and
Stockholm University, 12 Hannes Alfv\'ens v\"ag
10691 Stockholm, Sweden}
\date{\today}
\begin{abstract}
  We study a single, freely--floating, inextensible, elastic filament in a
  linear shear flow:
  $\UU_{0}(x,y) = \gdot y \xhat$.
  In our model: the elastic energy depends only on bending;  
  the rate--of--strain, $\gdot = S \sin(\omega t)$ is a periodic
  function of time, $t$; and 
  the interaction between the filament and the flow is approximated by
  a local isotropic drag force.
  Based on the shape of the filament we find five different
  dynamical phases: straight, buckled, periodic (with
  period two, period three, period four, etc), chaotic and one with
  chaotic transients.
  In the chaotic phase, we show that the iterative map for the angle,
  which the end--to--end vector
  of the filament makes with the tangent its one end,
  has period three solutions; hence it is chaotic. 
  Furthermore, in the chaotic phase the flow is an efficient mixer. 
\end{abstract}
\maketitle
\section{Introduction}
The dynamics of flexible filaments in flows plays a crucial role in
many biological and industrial processes~\cite{duprat2022moisture}.
A canonical example is that of cilia and flagella~\cite{brennen1977fluid,
  sleigh2016biology} that takes part in
wide variety of biological tasks, e.g., swimming of microorganisms,
feeding and breathing of marine invertebrates.
In such cases, although the flow nonlinearities can often be safely
ignored, due to its elastic nonlinearities and flow--structure
interactions a single isolated filament can show surprisingly complex dynamics
in flows.
Both active and passive filament, anchored or freely floating, in
various steady flows have been studied
extensively, see Ref.~\cite{du2019dynamics,bruot2016realizing} and
references therein.
In steady flows, a single passive filament has quite complex
transient dynamics~\cite{becker2001instability,
guglielmini2012buckling,liu2018morphological,lagrone2019complex,
slowicka2019flexible,zuk2021universal,kuei2015dynamics,hu2021levy,
chakrabarti2020flexible}.
For active filaments, the focus has been on how a periodic driving
can give rise to symmetry breaking, e.g., swimming~\cite{wiggins1998trapping}
or whirling~\cite{wolgemuth2000twirling,lim2004simulations,wada2006non}.
This year, three papers have focussed on, how periodic
driving, either of the flow or the filament, can give rise
to  secondary instabilities~\cite{bonacci2022dynamics} or statistically
stationary state with chaotic/complex
dynamics~\cite{agrawal2022chaos,krishnamurthy2022emergent}.
For the latter,
the shape of the filament, as described by its curvature as a function of
its arc length, is a spatiotemporally chaotic function.
Henceforth we call this phenomena \textit{shape chaos}. 
Such chaotic solutions are particularly interesting because they have the
potential to be used to generate efficient mixing in microfluidics. 

\begin{figure}
    \includegraphics[width=\linewidth]{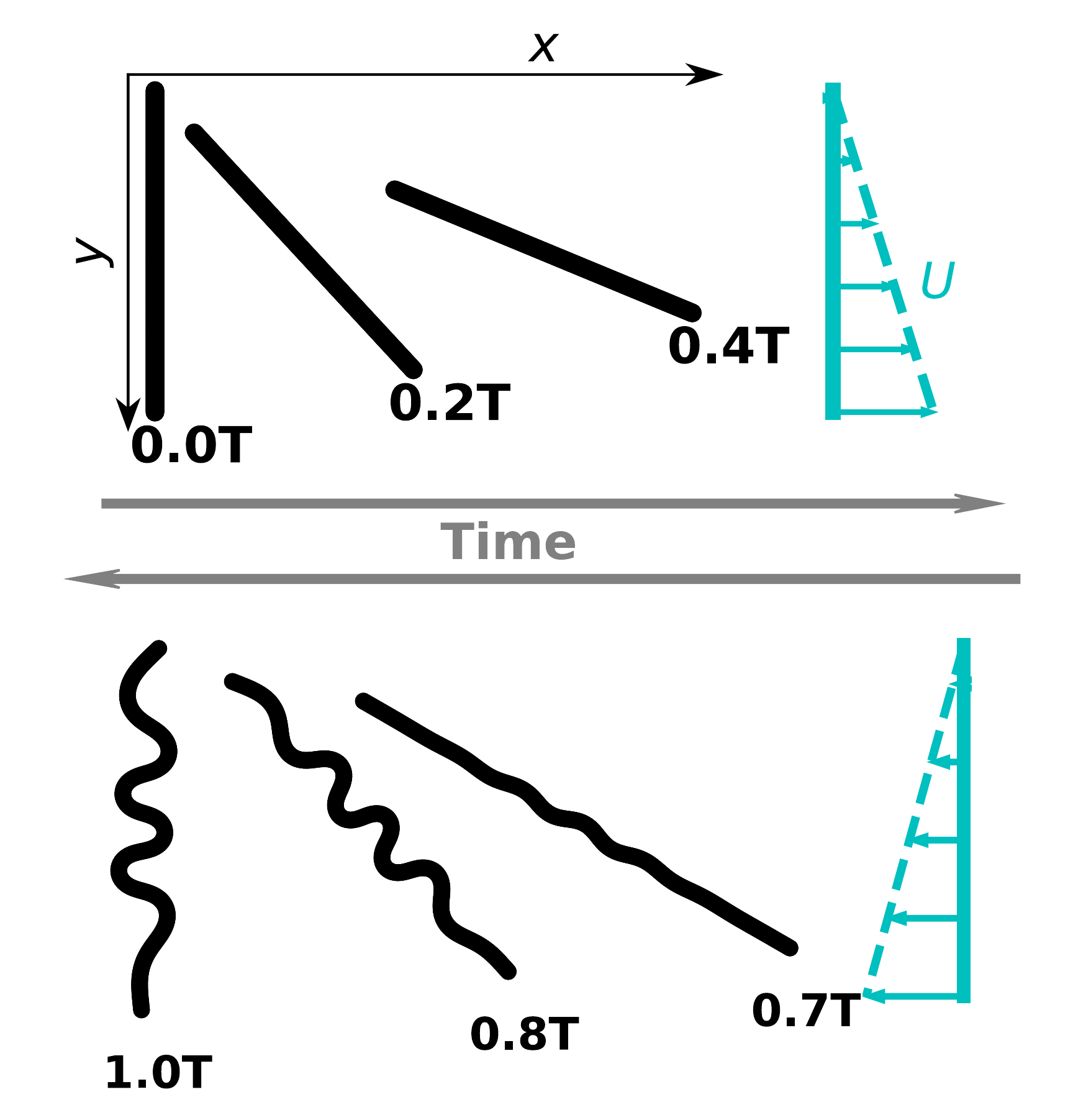}
    \caption{\textbf{Sketch of our numerical experiment}.
      Initially the filament is straight and is aligned vertically.
      The background shear flow, \Eq{eq:shear} is shown as arrow:
      $t=T/4$ (top panel) and $t=3T/4$ (bottom panel).
    \label{fig:snapshot}
    }
\end{figure}
Two effects determine the fate of an elastic filament in flow.
One is the elastic nonlinearity of the filament and the other
is the viscous interaction between the filament and the flow.
The latter, in all its glory, gives rise to non--local
and nonlinear interaction between
two different parts of the same filament.
Nevertheless, theoretical studies~\cite{Doi1986theory,goldstein1995nonlinear,
  goldstein1998viscous,wolgemuth2000twirling} have often
approximated the viscous effect as a local, linear, isotropic drag. 
Can this local approximation to the
flow--structure interaction still capture the shape chaos
of a freely-floating filament ?
As we show in the rest of this paper, the answer is yes;
we prove shape chaos using Sharkovskii
and Li and Yorke's famous result~\cite{Alligood1996chaos} --
existence of period orbits of period three
implies not only the existence of orbits of all periods
but also senstive dependence on initial condition.

In \Fig{fig:snapshot} we show a sketch of our numerical experiment.
Initially the filament is aligned vertically. The background
shear flow is given by
\begin{equation}
  \UU_{0}(x,y) = \gdot y \xhat\/,\quad\text{and}\quad
  \gdot = S \sin(\omega t)\/\label{eq:shear}.
\end{equation}
Here $T = 2\pi/\omega$ is the time period of the periodic shear and $S$ is
a constant. 
\section{Model}
We model the filament using the bead-spring model~\cite{larson1999brownian,
  guglielmini2012buckling, nazockdast2017fast,slowicka2019flexible,
  wada2006non,zuk2021universal}:
identical spherical beads of diameter $d$ are connected by over-damped
springs of equilibrium length $a$.
The position of the center of the $i$-th bead is $\RRi$,
where $i = 1 \ldots N$, the total number of beads.
The equation of motion is:
\begin{equation}
    \frac{\partial \Rat_\ri}{\partial t} =
    -\frac{1}{3\pi\eta d}\frac{\partial \hamil}{\partial \Rat_\ri} 
    + U_0^\alpha(\RRi)\/,
    \label{eq:dRdt}
\end{equation}
  where $U_0$ is given in \eq{eq:shear}. 
  Here $\eta$ is viscosity of the fluid,
  $\partial (\cdot)/\partial(\cdot)$ denotes partial derivative,
  $\UU_0$ is the velocity of the background shear,
  and $\mH$ is the elastic Hamiltonian of the filament.
  The Greek indices run from 1 to $D$, the dimensionality of the space, and
  the Latin indices run from 1 to $N$.
  The elastic Hamiltonian~\cite{wada2006non,wada2007stretching},
  has contributions from bending ($\mH_\rB$) and stretching ($\mH_\rS$):
  \begin{subequations}
    \begin{align}
      \hamil &= \mH_\rB + \mH_\rS \quad\text{where} \label{eq:H}\\
    \mH_{\rm B} &= aB\sum_{\ri=0}^{N-1} \kappa_{\ri}^2  
   \quad\text{and}
      \label{eq:bendingenergy} \\
      \mH_\rS &= \frac{H}{2a}
      \sum_{\ri=0}^{N-1} \left(\lvert \RR_{\ri+1} - \RR_\ri \rvert - a\right)^2\/;
      \quad\text{where}
      \label{eq:HS}\\
      \kappa_\ri &=  \frac{1}{a}\lvert\uuh_\ri\times\uuh_{\ri-1}\rvert\/\quad
      \text{and}
      \label{eq:kappa}\\
      \uuh_\ri &= \frac{\RR_{\ri+1} - \RR_\ri}{\lvert \RR_{\ri+1} -
         \RR_\ri \rvert}\/. 
  \end{align}
  \end{subequations}
Here $B$ is the bending modulus of the filament and $H$ is its stretching modulus. 
We ignore thermal fluctuations and torsion. 
Three dimensionless parameters determine the dynamics.
We call them, the elasto--viscous parameters, the dimensionless
frequency and the ratio of stretching to bending defined
respectively as:
\begin{subequations}
  \begin{align}
\mub &\equiv \frac{8\pi\eta S L^4}{B}\/, \label{eq:mu}\\
\sigma &\equiv \frac{\omega}{S}\/,\quad\text{and} \label{eq:sigma}\\
K &\equiv \frac{H a^2}{B}\/.
  \end{align}
\label{eq:param}
\end{subequations}
In practice, the filaments are  inextensible~\cite{powers2010dynamics},
which we implement by choosing appropriately high
value of $K$.
We evolve ~\Eq{eq:dRdt} using adaptive Runge-Kutta ~\cite{press1992adaptive}
method with cash-karp parameters ~\cite{cash1990variable}.
Our code is freely available~\footnote{
  \url{https://github.com/dhrubaditya/ElasticString}} and has been
benchmarked against experimental results~\cite{agrawal2022chaos}.
A complete list of the parameters of the simulation is given in
table~\ref{tab:parameters}.
We study the problem for a large range of $\mub$ and $\sigma$
all within experimentally realizable range.
Note that, with the local approximation of
viscous forces it is possible for the filament to cross itself.
Such unphysical solutions do appear in our simulations but for
values of $\mub$ other than that has been considered in this paper.
The computational complexity of the model, where the viscous
interaction is modelled by the non--local
Rotne--Pregor tensor~\cite{agrawal2022chaos},
is $\Ord(N^2)$ where $N$ is the number of beads, whereas
the computational complexity of the model with local viscosity is
$\Ord(N)$.
This allows us to run our simulations for much longer times than it
was possible in Ref.~\cite{agrawal2022chaos}.
\section{Results}
\begin{figure}[h]
    \includegraphics[width=\columnwidth]{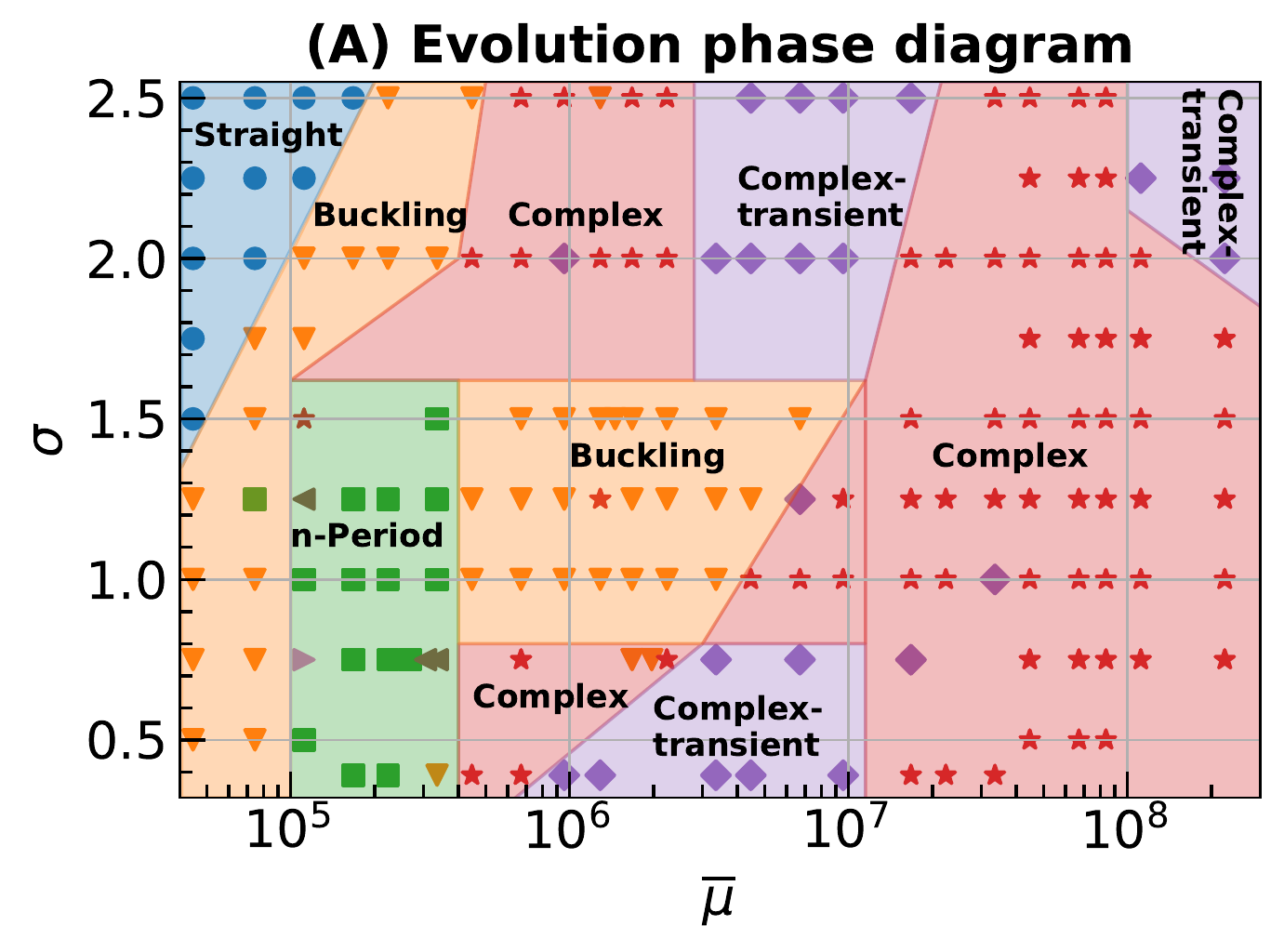}
    \includegraphics[width=\columnwidth]{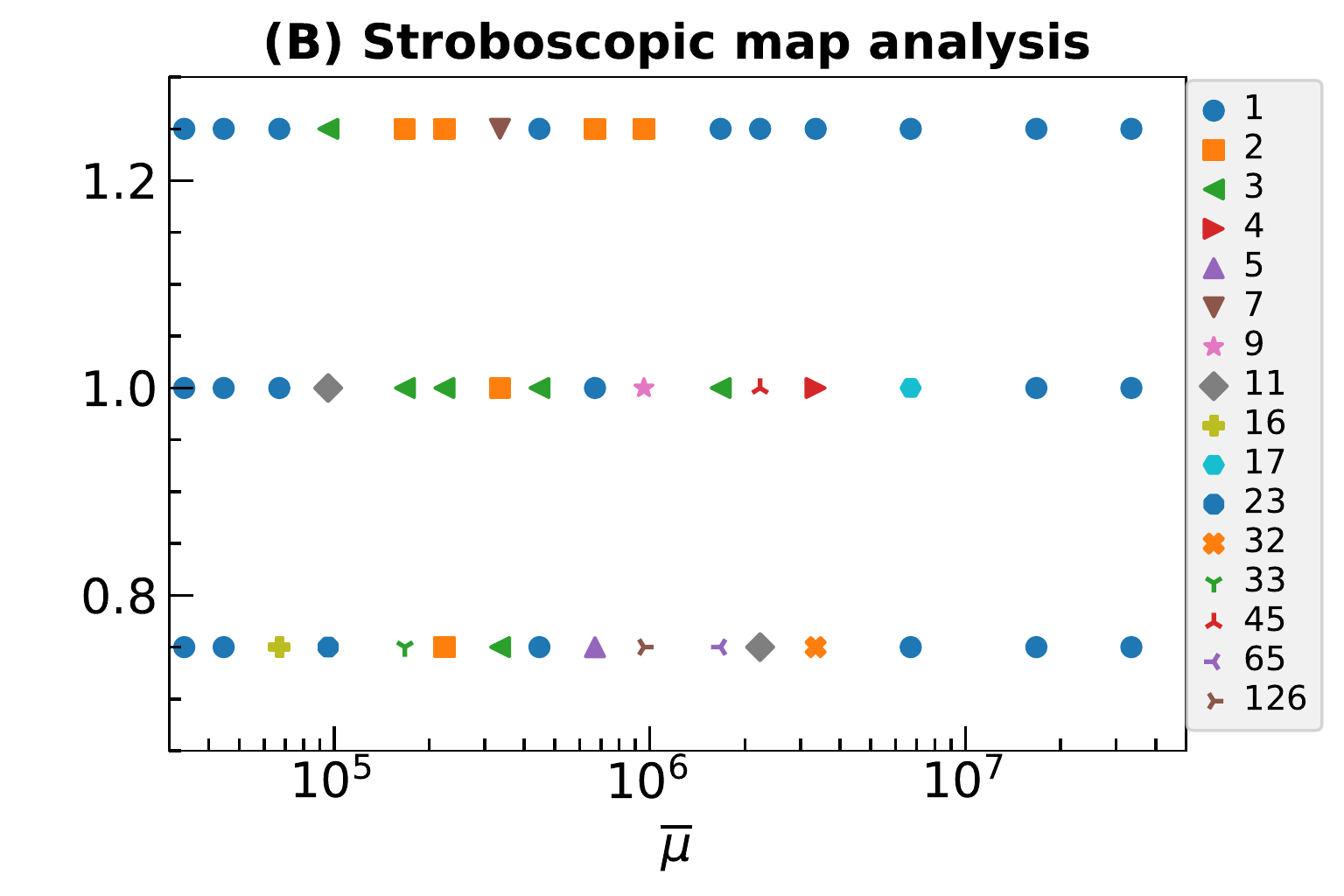}
    \caption{\textbf{(A):Phase diagram} in the $\mub$--$\sigma$ plane;
      We find $5$ different qualitatively different dynamical phases: 
      Straight($\bullet$); periodic ($\blacktriangledown$) 
      with n-period, where n=2($\blacksquare$), 3($\blacktriangleleft$),
      4($\blacktriangleright$); complex ($\bigstar$)
      complex-transients ($\blacklozenge$)
      \textbf{(B) Solutions of stroboscopic map}:
      The stroboscopic map has many periodic solutions at every point
      in $\mub$--$\sigma$ plane. We show time period of the lowest cycle in
      Sharvoskii ordering.
      \label{fig:phase}
    }
\end{figure}
A rigid ellipsoid in a periodic shear may show chaotic three--dimensional
rotation under certain conditions~\cite{ramamohan1994chaotic, kumar1995chaotic,
  lundell2011effect,nilsen2013chaotic}.
Such behavior emerges due to the nonlinearities present in the Euler's
equations of rigid body rotation.
Here we consider a filament with no inertia, hence such chaotic
solutions are not present in our system.
For a filament with high bending rigidity (small $\mub$)
we find that the filament merely translates and rotates
coming back to its initial position and shape at the end of
every period.

For a fixed dimensionless frequency ($\sigma$) as the bending rigidity
is decreased ($\mub$ is increased) an kaleidoscope of dynamic
behavior emerges.
We show an example in \Fig{fig:snapshot}.
During the first half-period the flow is extensional and the filament
rotates.
In the second half the flow is compressional and the filament
can undergo buckling transition -- 
the shape of the filament after one period is no longer straight
but buckled.
Furthermore, it may not come back to its initial position but
may come back translated or rotated, neither of which
are of interest to us in this paper -- we focus
on the shape of the filament. 
Under subsequent iterations of the periodic shear the
buckled filament can go through many changes in shape.

We show a dynamic phase diagram in \subfig{fig:phase}{A}.
Overall, at late times, the following possibilities exists:
\begin{enumerate}
\item The filament is always straight.
\item The filament reaches the same buckled shape at the end
  of each period.
\item The shape of the filament shows periodic behaviour
  with two-cycle, three-cycle, four-cycle, etc.
\item The shape of the filament is spatiotemporally chaotic.
  In \subfig{fig:phase}{A} such solutions are marked
  \textit{complex}. 
\item The filament shows chaotic behavior for a long time
  but such behavior turns out to be transient.
  At late times the filament settles down to a complicated
  shape which changes very slowly.
    In \subfig{fig:phase}{A} such solutions are marked
  \textit{complex transients}. 
\end{enumerate}
We have observed the same  qualitative behavior before~\cite{agrawal2022chaos},
for the case where the viscous forces are modeled by
the non-local Rotne-Pregor tensor,
with two crucial quantitative differences.
We did not observe any three--period solution before and
for large $\mub$ we obtained complex transients
for all values of $\sigma$ we used 
whereas here we observe the reappearance of the complex phase
for the higher $\mub$s.
Nevertheless, we conclude that the model with local viscosity is
able to capture the feature of the problem we consider essential
-- a rich dynamical phase diagram that includes complex shapes.

\subsection{Stroboscopic map}
The dynamical system described by \eq{eq:dRdt} is non-autonomous
because $\gdot$ is an explicit function of time.
Integrating \eq{eq:dRdt} over exactly one time period $T$
gives us the position of every bead of the filament at $t=T$.
Recall that the shape of the filament is fully specified by
its curvature $\kappa$ as a function of arc length $s$.
Thus we can define the stroboscopic map, $\cF$, that allows us
to obtain $\kappa(s,T)$ from $\kappa(s,0)$: 
\begin{equation}
\kappa(s,T) = \cF \kappa(s,0)\/.
\label{eq:map}
\end{equation}
The stroboscopic map is no longer an explicit function of time.
Following Refs.~\cite{auerbach1987exploring, cvitanovic2005chaos},
we study the shape-chaos by obtaining the fixed points
and periodic orbits of the stroboscopic map using the Newton--Krylov
method, which is described in detail in appendix D1 of our earlier
paper~\cite{agrawal2022chaos}.
In general, for any fixed value of $\mub$ and $\sigma$ we obtain
many periodic orbits.
We list all of them in table~\ref{tab:cycles}.
We sort the cycles using \sharv's ordering~\cite{sharkovskiui1995coexistence}:
\begin{eqnarray}
  3 &\prec& 5 \prec 7 \prec 9 \prec 11 \ldots \prec 2\cdot 3 \prec 2\cdot5 \prec\ldots \nonumber \\
  \ldots &\prec& 2^2\cdot 3 \prec 2^2\cdot 5 \ldots \prec 2^3\cdot 3 \prec 2^3\cdot 5\prec \ldots
  \nonumber \\
  \ldots &\prec& 2^3\prec2^2\prec 2 \prec1\/,
  \label{eq:shar}
\end{eqnarray}
In \subfig{fig:phase}{B} we show the leading period of
stroboscopic map, as it appears in Sharkovskii's ordering,
as a function of $\mub$ and $\sigma$ -- we do
find orbits of period three. 
Although many periodic orbits appear as solutions of the map most of them
are not stable and do not appear in the solution of dynamical
equation.
Let us recall the \sharv's theorem~\cite{Alligood1996chaos}:
Consider a continuous  map $f$ on an interval
with a period $p$ orbit.
If $p \prec q$, where $q$ appears in the \sharv's ordering,
then $f$ has a period-$q$ orbit.
This implies that if $f$ possess
a period $3$ orbit it has all orbits of all other periods.
Although this shows that the map has very complex dynamical behavior
it does not necessarily proves the existence of chaos.
Nevertheless, existence of period three does imply chaos
as was proved by Li and Yorke~\cite{li2004period}.
Unfortunatley neither \sharv's theorem nor the result of Li and Young
is valid for maps in dimensions higher than unity\footnote{
  As a counterexample~\cite{kloeden2006li}, consider the two dimensional map
  that rotates every point in the $x-y$ plane by an angle of $2\pi/3$
  in the counter-clockwise direction. 
  Clearly this map has a period three solution but
  it is not chaotic.}
Hence we conclude that although we demonstrate the rich complexity of the
solutions of the stroboscopic map and we have not yet conclusively proven the
existence of chaotic solutions.

\subsection{Period three and the $\Theta$ map}
\begin{figure*}
    \includegraphics[width=0.32\linewidth,height=0.3\linewidth]{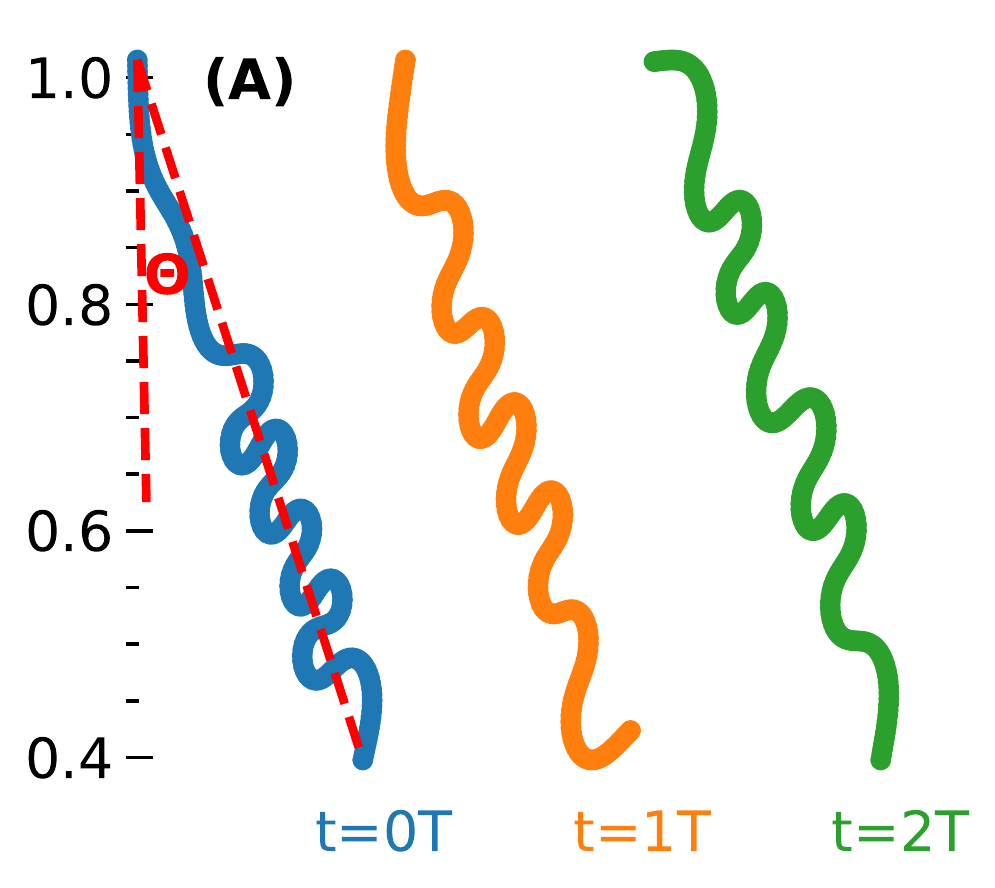}
    \hspace*{-10pt}
    \includegraphics[width=0.48\linewidth,height=0.30\linewidth]{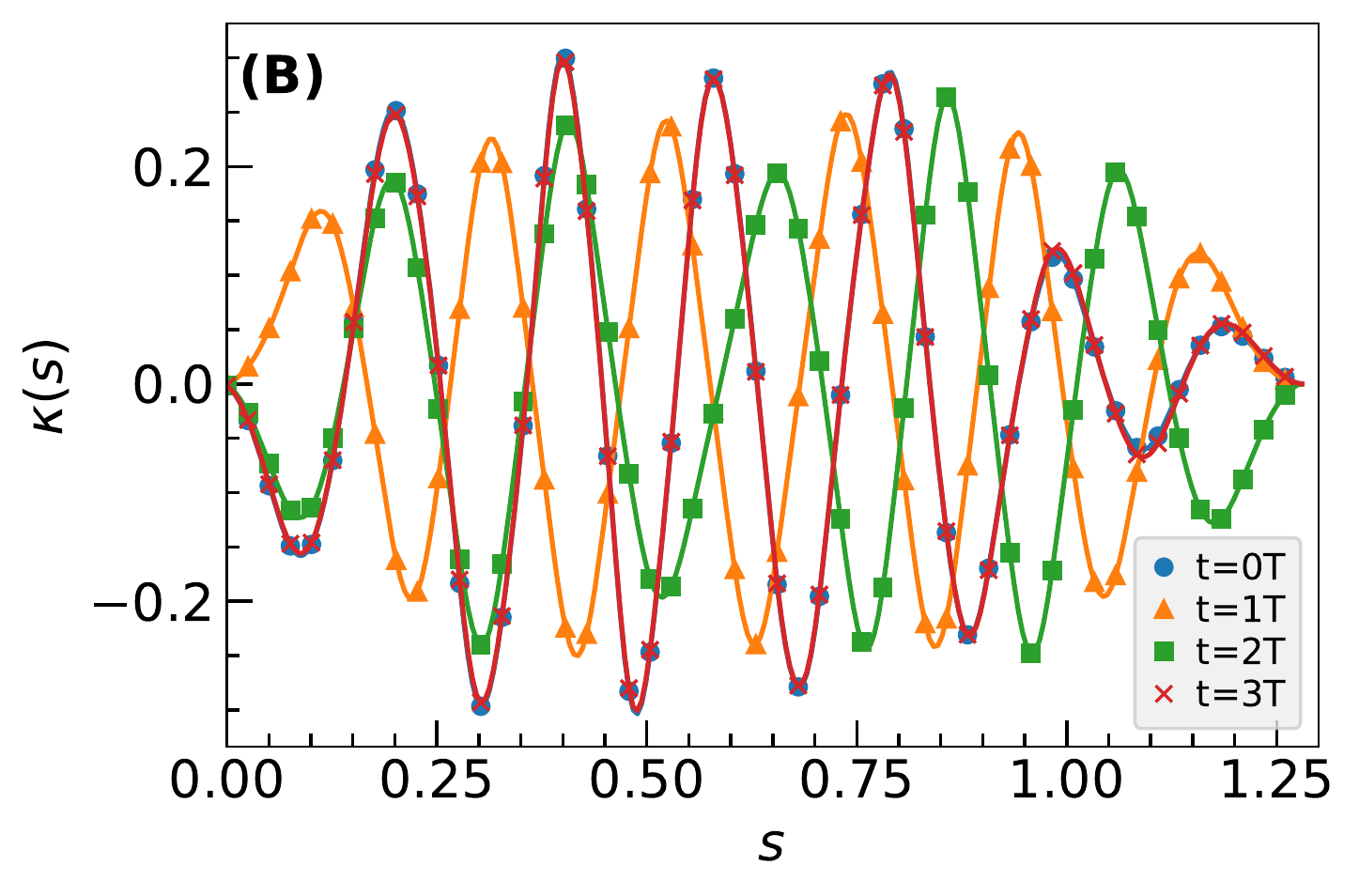}
    \hspace*{-10pt}
    \includegraphics[width=0.21\linewidth,height=0.30\linewidth]{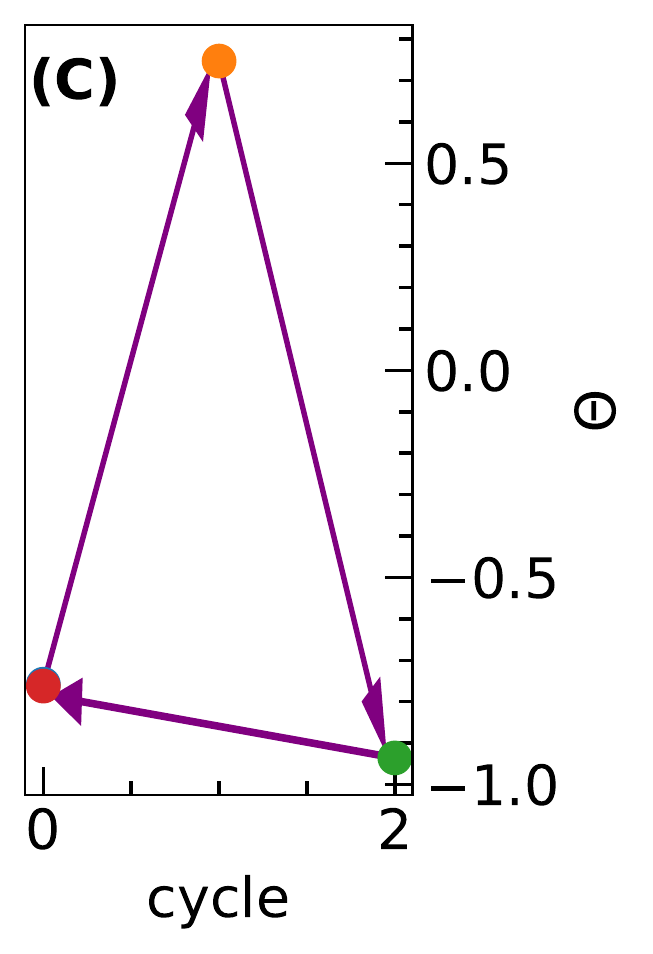}
    \caption{{\bf Example of a three period solution}
      (for $\sigma=1,\mub=1.67\times10^6$) in real space (A) and
      curvature space (B) and for ($\sigma=0.75,\mub=3.35\times10^5$) in
      1D $\Theta$-space (C).
      We start from $t=0$ and evolve the filament for three cycles.
      In (A,B), blue, orange and green curves shows the filament at
      $t=0,1T$, and $2T$ respectively.
      In (B), the $\kappa(s)$ plots 
      for $t=0$ and $3T$ (blue, red) lie on top of each other.
    \label{fig:threep}
    }
\end{figure*}
\begin{figure*}
  \includegraphics[width=\linewidth]{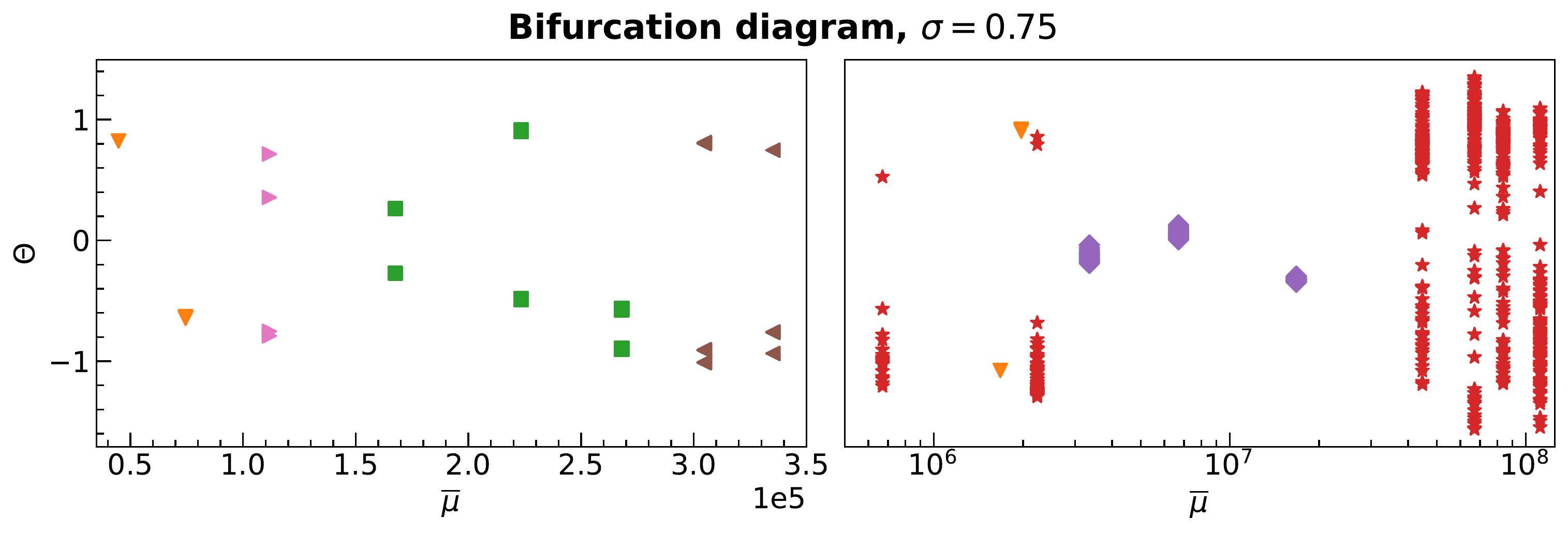}
  \caption{\textbf{Bifurcation diagram} of $\Theta$ for $\sigma=0.75$.
      The system shows chaos at high enough $\mub$, except certain isolated
      ranges of $\mub$, where $\Theta$ settles into periodic behavior --
      a behavior similar to island of order for bifurcation diagram of
      the logistic map~\cite{Alligood1996chaos}.}
  \label{fig:Theta}
\end{figure*}
The existence of period three solutions for both the
evolution equations and the stroboscopic map is a key result in this problem.
It behooves us to study it in greater detail.
We choose the period-three solution that appears as
a solution of stroboscopic map for $\sigma=1$ and $\mub = 1.6\times 10^6$.
This is shown in \subfig{fig:phase}{B} as a green triangle.
In \subfig{fig:threep}{A,B} we show the three solutions in real and curvature space respectively.

Now we attempt an arbitrary dimensional reduction to construct a
one-dimensional map.
We draw the straight line that connects the top end of the filament to
the bottom one and call $\Theta$ the angle this line makes with the
tangent to the filament at its top point, see \subfig{fig:threep}{A}.
Defined this way, $\Theta$ does not depend on the position or orientation
of the overall filament, but only its
shape.
In some cases, if shape of the filament shows a period three solution,
so does $\Theta$. 
One such case is shown in~\subfig{fig:threep}{C} for $\sigma=0.75,\mub=3.35\times10^5$~\footnote{
It is also possible that the shape shows period three solution but the 
$\Theta$ shows a fixed point or a period--two solution.
Conversely, it is possible for $\Theta$ to have a period--three
solution without the shape having a period-three solution.}.
From the stroboscopic map we can construct a map for $\Theta$.
This is a one dimensional map for which the Li and Yorke theorem
is valid.
Thus by demonstrating that the $\Theta$ map has three period we
show that this map is chaotic.
Further evidence of chaos is obtained by plotting the
bifurcation diagram for $\Theta$ in \fig{fig:Theta}.
We find period two, period four and period three solutions
and also chaotic ones. 

\subsection{Mixing of passive tracers}
\begin{figure}
  \includegraphics[width=0.49\columnwidth]{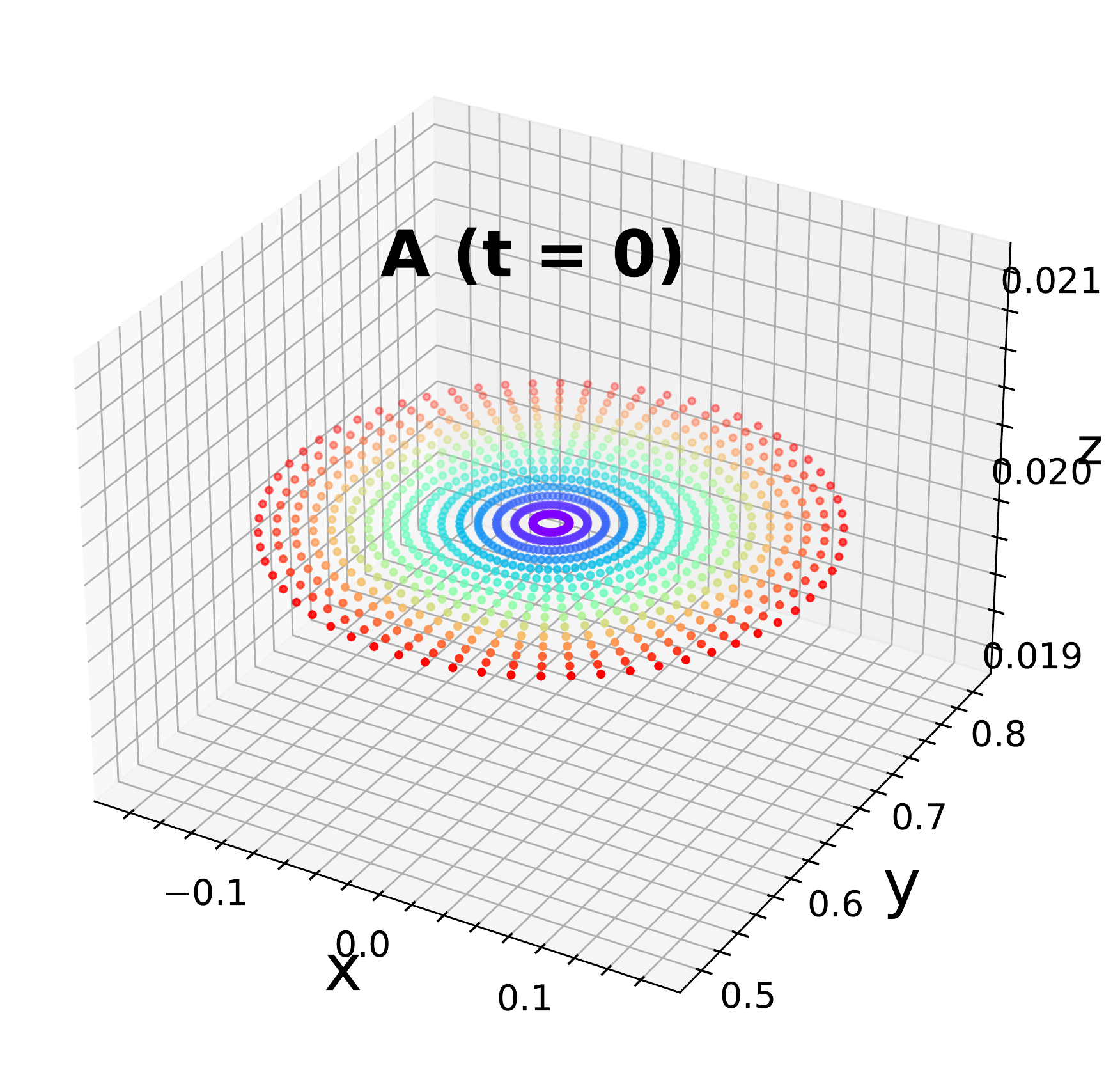}
  \includegraphics[width=0.49\linewidth]{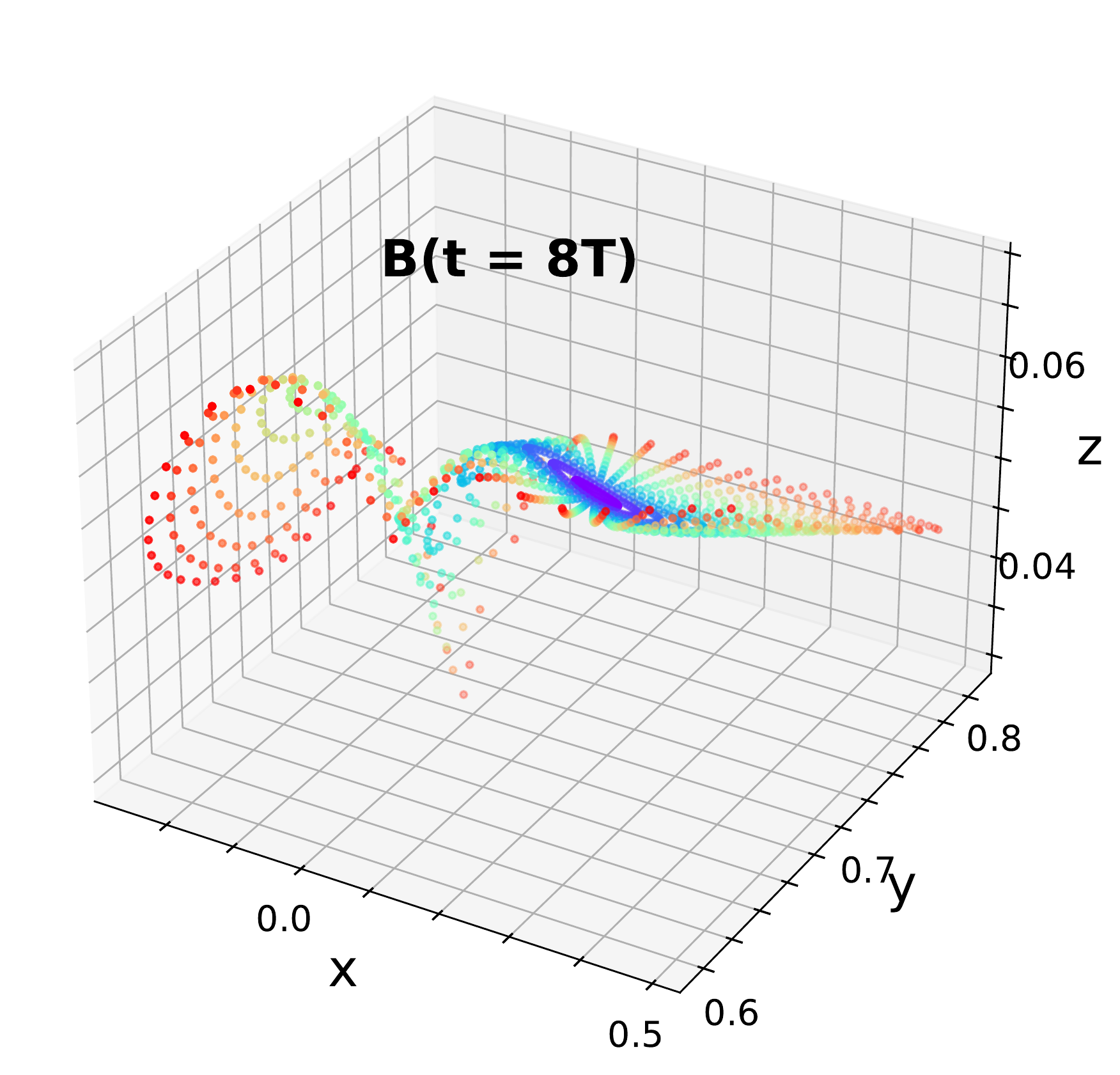} \\
  \includegraphics[width=\linewidth]{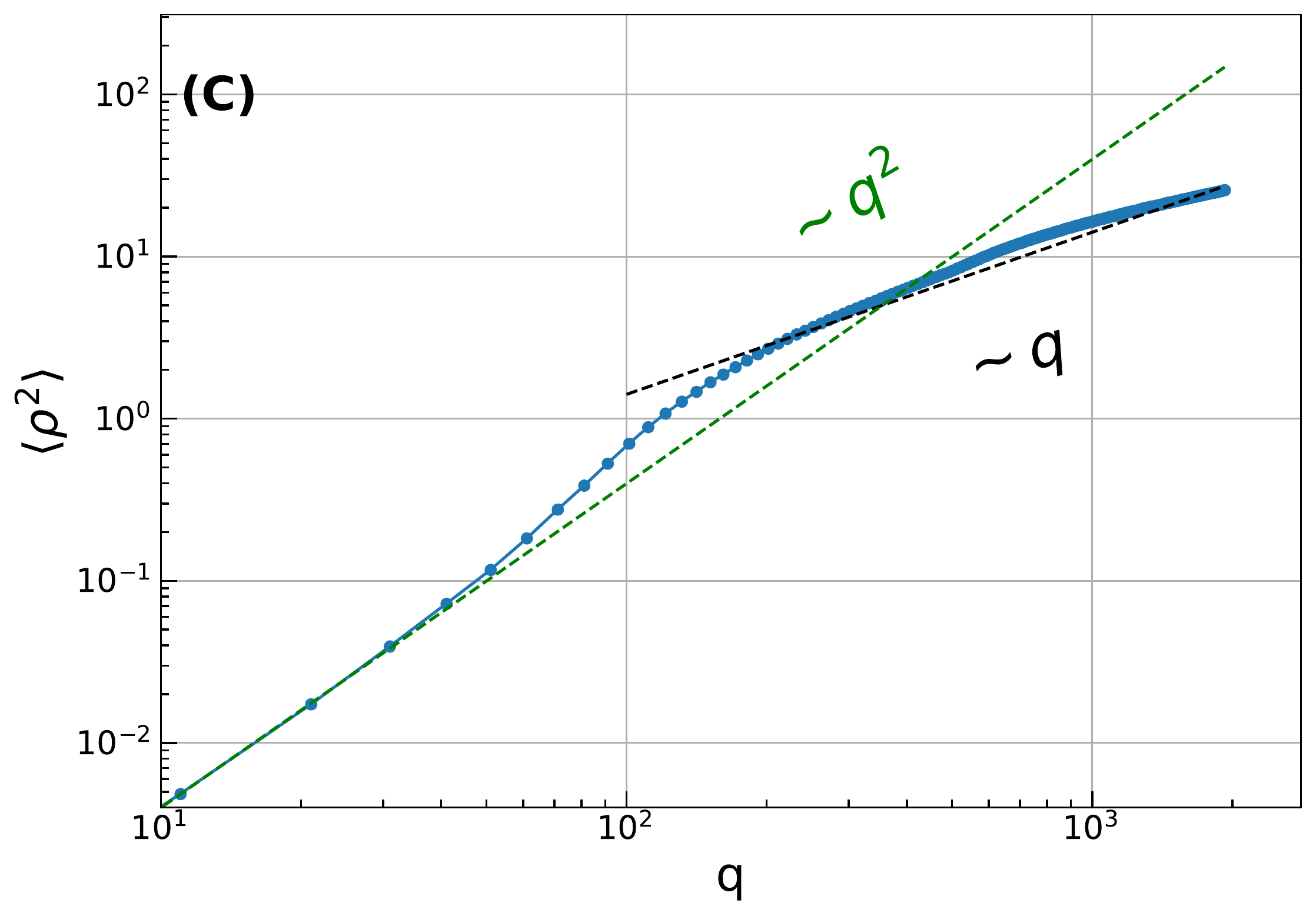} \\
  \caption{\textbf{Mixing of passive tracer}:
    (A) and (B)  Positions of tracer particles  
    at $t=0$ and $t=8T$.
    The parameters  $\sigma=0.75$ and $\mub=6.7\times10^7$
    are chosen such that we are in the complex phase. 
    Initially, the tracers are placed on concentric circles, color coded by
    their distances from the center of the circles.
    The mixing of the colors show the mixing of the tracers. 
    (C) Mean square displacement (MSD), $\bra{\rho^2}(qT)$, defined in
    \eq{eq:msd},  as a function of $q$ in log-log scale.
    We also plot two lines with slopes $1$ and $2$. 
   \label{fig:mixing} }
\end{figure}
Next we demonstrate that if we choose $\mub$ and $\sigma$ inside the
complex phase, see \subfig{fig:phase}{A}, then the filament acts as
an effective mixer of passive tracers.
We use the same notation and technique used in our earlier
paper~\cite{agrawal2022chaos}.

Once the filament has reached a statistically stationary
state we introduce $\Np$ tracers placed on  concentric circles
in the $x$--$y$ plane, \subfig{fig:mixing}{A}.
They are colored by radius of the circle on which they lie on at the initial
time.
For the rest of this section the time when the tracers
are introduced is $t=0$. 
The equation of motion of the k-th tracer particle
whose position at time $t$ is given
by $\XX^{\rk}(t)$, is
\begin{equation}
  \frac{d\XX^{\rk}}{dt} = \UU(\rr)\delta(\rr-\XX^{\rk})\/.
  \label{eq:tracer}
\end{equation}
Here $\UU(\rr)$, the velocity of the flow 
at  $\rr = (x,y,z)$,
is a superposition of the background flow velocity $\UU_0(\rr)$
and the contributions from all the beads in the
filament~\cite{rotne1969variational,
  brady1988stokesian, guazzelli2011physical,kim2013microhydrodynamics}:
\begin{subequations}
  \begin{align}
    U^{\alpha}(\rr) &= U_0^{\alpha}(\rr) + \mG^{\ab}(\rr-\RRi)F^{\beta}_{\rm i}\/,\quad\/,
      \text{where}\label{eq:Ueuler}\\
      F^{\alpha}_{\rm i} &= -\frac{\partial \hamil}{\partial R^{\alpha}_{\rm i}}\/,
      \quad\text{and}\label{eq:ff}\\
      \mG^{\ab}(\RR) &= \frac{1}{8\pi\eta R}\left[ \dab + \frac{\Ra\Rb}{R^2}
        +\frac{d^2}{4R^2}\left( \frac{1}{3}\dab - \frac{\Ra\Rb}{R^2} \right) \right]
      \label{eq:Green}
  \end{align}
  \label{eq:tracervel}
\end{subequations}
At $t=8T$, we find that most of the tracers have moved out of the plane
and have become somewhat mixed, \subfig{fig:mixing}{B}.  
At even later time, $t=128T$ (not shown), we find the tracer particles are
well mixed.
To obtain a quantitative measure of mixing we define
\begin{equation}
  \Delta \XX^{\rm k}_{\rm j} \equiv \XX^{\rm k}((j+1)T) - \XX^{\rm k}(jT)\quad\/,
  \label{eq:Dx}
\end{equation}
the net displacement of the $\rk$-th tracer particle over the
$j$-th cycle -- $t=jT$ to $t=(j+1)T$, where $j$ is
an integer.
The net displacement of the $\rk$-th tracer after $q$ cycles is
\begin{equation}
  \brho^{\rk}(q) = \sum_{j=1}^{q}\Delta \XX^{\rk}_{\rj}\/.
  \label{eq:rho}
\end{equation}
The total mean square displacement, averaged over all the tracers, at the end of
$q$ cycles is given by
\begin{equation}
  \bra{\rho^2(q)} \equiv \frac{1}{\Np}\sum_{\rk=1}^{\Np}
  \lvert \brho^{\rk}(q) \rvert^2 \label{eq:msd} 
\end{equation}
In \subfig{fig:mixing}{C} we plot $\bra{\rho^2(q)}$ versus $q$ in log-log
scale.
If the tracers diffuse then we expect $\bra{\rho^2(q)} \sim q$ for
large $q$~\cite{taylor1922diffusion}, which is what we obtain.
Furthermore, we calculate the cumulative probability density function (CPDF)
for each component of the displacement $\Delta \XX^{\rk}_{\rj}$.
For the out-of-plane component this CPDF has an exponential tail.
For the in-plane components we obtain a power-law tail of exponent of $-3$.
This implies that the probability density function (PDF) of each component
of the displacement $\Delta \XX^{\rk}_{\rj}$ is such
that its second moment is well defined.
Hence by the central limit theorem the probability density function of
$\rho^2(q)$ is a Gaussian and we expect simple diffusive behavior. 
However, as the PDF (of displacement) has power-law tail we expect that very
long averaging
over very many number of tracer particles is necessary for convergence.
This explains why we observe not-so-clear evidence of diffusion.
\section{Conclusion}
In this paper we consider a simplified model for a flexible filament in
a viscous flow driven in a time--periodic manner.
In particular, the simplicity lies is approximating the viscous forces
by a local drag.
We show that the shape of the filament is spatiotemporally chaotic.
This model has only the elastic nonlinearity of the filament, hence
it is solely the elastic nonlinearity that is responsible for
chaos. This is the central message of this paper.
An additional advantage of using the simplified model for viscous
forces is that it may be possible to make theoretical progress
following Goldstein and Langer~\cite{goldstein1995nonlinear}. 

The dimensionless parameters that we consider are within a range that
is experimentally accessible. 
Although we do not expect exact
quantitative agreement with experiment,
We hope that, together
with our previous work~\cite{agrawal2022chaos}, we have now presented 
convincing
evidences that a single flexible filament in periodically driven
Stokes flow can give chaotic solutions that is able to effectively
mix passive scalars even at infinite Peclet number. 
\begin{table}
\centering
\begin{ruledtabular}
\begin{tabular}{ cc }
$N$ & $256$  \\[0.5em] 
$a$ & $0.005$  \\[0.5em] 
$d$ & $0.005$\\[0.5em]
$L$ & $1.28$\\[0.5em]
$B$ & $6\times10^{-6} \mbox{--} 4\times10^{-2}$\\[0.5em]
$S$  & $2$ \\[0.5em]
$\eta$ & $10$\\[0.5em]
$\omega$ & $1 \mbox{--} 6$ \\[0.5em]
$\Delta$ & $10^{-4} \mbox{--} 10^{-6}$ \\[0.5em]
$\mub$ & $12560 \mbox{--} 8.37 \times 10^7 $ \\[0.5em]
$\sigma$ & $0.5 \mbox{--} 3$ \\[0.5em]
$K$ & $16$\\
\end{tabular}
\caption{Parameters of simulation:
  Number of beads, $N$,
  equilibrium distance between beads, $a$,
  bead diameter $d$,
filament length $L$ 
Bending modulus $B$,
viscosity $\eta$,
time-step, $\Delta $.
The quantities $S$ and $\omega$ defined in \eq{eq:shear}
determined the space-time dependence of the background flow.
The dimensionless parameters $\mub$, $\sigma$ and $K$
are defined in \eq{eq:param}. 
}
\label{tab:parameters}
\end{ruledtabular}
\end{table}
\begin{table*}
\centering
\begin{ruledtabular}
\begin{tabular}{c  c  c  c}
\backslashbox{$\mub$}{$\sigma$} & 0.75 &  1.0 & 1.25 \\
\hline
$3.35\times10^4$ & 1 & 1 & 1 \\[0.5em]
$4.47\times10^4$ & 1 & 1 & 1 \\[0.5em]
$6.67\times10^4$ & 1,16 & 1 & 1 \\[0.5em]
$9.57\times10^4$ & 1,4,20,22,23,27 & 2,4,11 & 1,2,3,12,13,24,25 \\[0.5em]
$1.67\times10^5$ & 1,2,10,14,33 & 1,2,3 & 1,2 \\[0.5em]
$2.23\times10^5$ & 1,2& 1,2,3& 1,2\\[0.5em]
$3.35\times10^5$ & 1,2,3,4,5,7,9,10,15,19,22,23,31 & 1,2& 1,2,7\\[0.5em]
$4.47\times10^5$ & 1 & 1,2,3,5,6,7,13,19,37,41 & 1\\[0.5em]
$6.70\times10^5$ & 1,2,5,7,12,25,32,41,47,50,68,76,85,100,104 & 1 & 1,2 \\[0.5em]
$9.57\times10^5$ & 1,126 & 1,4,9 & 1,2 \\[0.5em]
$1.67\times10^6$ & 1,50,54,62,65,66,70,80,86,108,132 & 1,2,3,4,5,7,9,22,44 & 1\\[0.5em]
$2.23\times10^6$ & 1,11,12,28 & 1,2,45 &1 \\[0.5em]
$3.34\times10^6$ & 1,2,4,32 & 1,4 & 1\\[0.5em]
$6.67\times10^6$ & $1^\ast$ & 1,17 & 1 \\[0.5em]
$1.67\times10^7$ & $1^\ast$ & $1^\ast$ & $1^\ast$\\[0.5em]
$3.35\times10^7$ & $1^\ast$ & $1^\ast$ & $1^\ast$\\[0.5em]
\end{tabular}
\caption{\textbf{Cycles of the stroboscopic map}.
  In some of these cases the evolution equation shows chaotic solution
  but so far we have obtained one or two periodic orbits of small periods,
  these are marked by $\ast$. }
\label{tab:cycles}
\end{ruledtabular}
\end{table*}
\input{paper.bbl}
\end{document}

%% file: symdef.inc
\def \Ord {\mathcal{O}}

\def \RRi {\bm{R}_{\rm i}}

\def \gdot {\dot{\gamma}}

\def \ri {\rm i}

\def \rr {\bm{r}}

\def \xhat {\hat{\rm  x}}

\def \XX {{\bm{X}}}

\def \ri {{\rm i}}
\def \rj {{\rm j}}
\def \rk {{\rm k}}

\def \del2z {\partial^{2}_{z}}

\def \dab {\delta_{\alpha\beta}}

\def \ab {{\alpha\beta}}

\def \uuh {\hat{\bm {u}}}

\def \UU {{\bm U}}

\newcommand{\eq}[1]{(\ref{#1})}

\newcommand{\Eq}[1]{Equation~(\ref{#1})}

{}

\def \RR {\bm{R}}
\newcommand{\fig}[1]{Fig.~\ref{#1}}
\newcommand{\subfig}[2]{Fig.~\ref{#1}(#2)}
\newcommand{\Fig}[1]{Figure.~\ref{#1}}

\def\drawing #1 #2 #3 {
\begin{center}
\setlength{\unitlength}{1mm}
\begin{picture}(#1,#2)(0,0)
\put(0,0){\framebox(#1,#2){#3}}
\end{picture}
\end{center} }

\def \mG {\mathcal{G}}

\def \mub {\overline{\mu}}

\def \mH {\mathcal{H}}

\def \Ra {R_\alpha}
\def \Rb {R_\beta}

\def \Rat {R^\alpha}

\def \hamil {\mathcal{H}}

\def \th {{\rm {\th}}}

\def \rk {{\rm k}}

\def \th {{\rm{th} }}

\def \rB {{\rm {B}}}
\def \rS {{\rm {S}}}

\def \cF 	{\mathcal{F}}
\def \Np {N_{\rm p}}
\def \brho {\bm{\rho}}
\newcommand{\bra}[1]{\left\langle #1\right\rangle}

\newcommand{\succprec}{\mathrel{\mathpalette\succ@prec{\succ\prec}}}
\newcommand{\precsucc}{\mathrel{\mathpalette\succ@prec{\prec\succ}}}
\def \sharv {Sharkovski\v{\i}}

%% file: paper.bbl
%

%% file: paper.bbl
\begin{thebibliography}{48}%
\makeatletter
\providecommand \@ifxundefined [1]{%
 \@ifx{#1\undefined}
}%
\providecommand \@ifnum [1]{%
 \ifnum #1\expandafter \@firstoftwo
 \else \expandafter \@secondoftwo
 \fi
}%
\providecommand \@ifx [1]{%
 \ifx #1\expandafter \@firstoftwo
 \else \expandafter \@secondoftwo
 \fi
}%
\providecommand \natexlab [1]{#1}%
\providecommand \enquote  [1]{``#1''}%
\providecommand \bibnamefont  [1]{#1}%
\providecommand \bibfnamefont [1]{#1}%
\providecommand \citenamefont [1]{#1}%
\providecommand \href@noop [0]{\@secondoftwo}%
\providecommand \href [0]{\begingroup \@sanitize@url \@href}%
\providecommand \@href[1]{\@@startlink{#1}\@@href}%
\providecommand \@@href[1]{\endgroup#1\@@endlink}%
\providecommand \@sanitize@url [0]{\catcode `\\12\catcode `\$12\catcode
  `\&12\catcode `\#12\catcode `\^12\catcode `\_12\catcode `\%12\relax}%
\providecommand \@@startlink[1]{}%
\providecommand \@@endlink[0]{}%
\providecommand \url  [0]{\begingroup\@sanitize@url \@url }%
\providecommand \@url [1]{\endgroup\@href {#1}{\urlprefix }}%
\providecommand \urlprefix  [0]{URL }%
\providecommand \Eprint [0]{\href }%
\providecommand \doibase [0]{http://dx.doi.org/}%
\providecommand \selectlanguage [0]{\@gobble}%
\providecommand \bibinfo  [0]{\@secondoftwo}%
\providecommand \bibfield  [0]{\@secondoftwo}%
\providecommand \translation [1]{[#1]}%
\providecommand \BibitemOpen [0]{}%
\providecommand \bibitemStop [0]{}%
\providecommand \bibitemNoStop [0]{.\EOS\space}%
\providecommand \EOS [0]{\spacefactor3000\relax}%
\providecommand \BibitemShut  [1]{\csname bibitem#1\endcsname}%
\let\auto@bib@innerbib\@empty
\bibitem [{\citenamefont {Duprat}(2022)}]{duprat2022moisture}%
  \BibitemOpen
  \bibfield  {author} {\bibinfo {author} {\bibfnamefont {C}~\bibnamefont
  {Duprat}},\ }\bibfield  {title} {\enquote {\bibinfo {title} {Moisture in
  textiles},}\ }\href@noop {} {\bibfield  {journal} {\bibinfo  {journal}
  {Annual Review of Fluid Mechanics}\ }\textbf {\bibinfo {volume} {54}},\
  \bibinfo {pages} {443--467} (\bibinfo {year} {2022})}\BibitemShut {NoStop}%
\bibitem [{\citenamefont {Brennen}\ and\ \citenamefont
  {Winet}(1977)}]{brennen1977fluid}%
  \BibitemOpen
  \bibfield  {author} {\bibinfo {author} {\bibfnamefont {Christopher}\
  \bibnamefont {Brennen}}\ and\ \bibinfo {author} {\bibfnamefont {Howard}\
  \bibnamefont {Winet}},\ }\bibfield  {title} {\enquote {\bibinfo {title}
  {Fluid mechanics of propulsion by cilia and flagella},}\ }\href@noop {}
  {\bibfield  {journal} {\bibinfo  {journal} {Annual Review of Fluid
  Mechanics}\ }\textbf {\bibinfo {volume} {9}},\ \bibinfo {pages} {339--398}
  (\bibinfo {year} {1977})}\BibitemShut {NoStop}%
\bibitem [{\citenamefont {Sleigh}(2016)}]{sleigh2016biology}%
  \BibitemOpen
  \bibfield  {author} {\bibinfo {author} {\bibfnamefont {Michael~A}\
  \bibnamefont {Sleigh}},\ }\href@noop {} {\emph {\bibinfo {title} {The biology
  of Cilia and Flagella: international series of monographs on pure and applied
  biology: zoology, vol. 12}}},\ Vol.~\bibinfo {volume} {12}\ (\bibinfo
  {publisher} {Elsevier},\ \bibinfo {year} {2016})\BibitemShut {NoStop}%
\bibitem [{\citenamefont {Du~Roure}\ \emph {et~al.}(2019)\citenamefont
  {Du~Roure}, \citenamefont {Lindner}, \citenamefont {Nazockdast},\ and\
  \citenamefont {Shelley}}]{du2019dynamics}%
  \BibitemOpen
  \bibfield  {author} {\bibinfo {author} {\bibfnamefont {Olivia}\ \bibnamefont
  {Du~Roure}}, \bibinfo {author} {\bibfnamefont {Anke}\ \bibnamefont
  {Lindner}}, \bibinfo {author} {\bibfnamefont {Ehssan~N}\ \bibnamefont
  {Nazockdast}}, \ and\ \bibinfo {author} {\bibfnamefont {Michael~J}\
  \bibnamefont {Shelley}},\ }\bibfield  {title} {\enquote {\bibinfo {title}
  {Dynamics of flexible fibers in viscous flows and fluids},}\ }\href@noop {}
  {\bibfield  {journal} {\bibinfo  {journal} {Annual Review of Fluid
  Mechanics}\ }\textbf {\bibinfo {volume} {51}},\ \bibinfo {pages} {539--572}
  (\bibinfo {year} {2019})}\BibitemShut {NoStop}%
\bibitem [{\citenamefont {Bruot}\ and\ \citenamefont
  {Cicuta}(2016)}]{bruot2016realizing}%
  \BibitemOpen
  \bibfield  {author} {\bibinfo {author} {\bibfnamefont {Nicolas}\ \bibnamefont
  {Bruot}}\ and\ \bibinfo {author} {\bibfnamefont {Pietro}\ \bibnamefont
  {Cicuta}},\ }\bibfield  {title} {\enquote {\bibinfo {title} {Realizing the
  physics of motile cilia synchronization with driven colloids},}\ }\href@noop
  {} {\bibfield  {journal} {\bibinfo  {journal} {Annual Review of Condensed
  Matter Physics}\ }\textbf {\bibinfo {volume} {7}},\ \bibinfo {pages}
  {323--348} (\bibinfo {year} {2016})}\BibitemShut {NoStop}%
\bibitem [{\citenamefont {Becker}\ and\ \citenamefont
  {Shelley}(2001)}]{becker2001instability}%
  \BibitemOpen
  \bibfield  {author} {\bibinfo {author} {\bibfnamefont {Leif~E}\ \bibnamefont
  {Becker}}\ and\ \bibinfo {author} {\bibfnamefont {Michael~J}\ \bibnamefont
  {Shelley}},\ }\bibfield  {title} {\enquote {\bibinfo {title} {Instability of
  elastic filaments in shear flow yields first-normal-stress differences},}\
  }\href@noop {} {\bibfield  {journal} {\bibinfo  {journal} {Physical Review
  Letters}\ }\textbf {\bibinfo {volume} {87}},\ \bibinfo {pages} {198301}
  (\bibinfo {year} {2001})}\BibitemShut {NoStop}%
\bibitem [{\citenamefont {Guglielmini}\ \emph {et~al.}(2012)\citenamefont
  {Guglielmini}, \citenamefont {Kushwaha}, \citenamefont {Shaqfeh},\ and\
  \citenamefont {Stone}}]{guglielmini2012buckling}%
  \BibitemOpen
  \bibfield  {author} {\bibinfo {author} {\bibfnamefont {Laura}\ \bibnamefont
  {Guglielmini}}, \bibinfo {author} {\bibfnamefont {Amit}\ \bibnamefont
  {Kushwaha}}, \bibinfo {author} {\bibfnamefont {Eric~SG}\ \bibnamefont
  {Shaqfeh}}, \ and\ \bibinfo {author} {\bibfnamefont {Howard~A}\ \bibnamefont
  {Stone}},\ }\bibfield  {title} {\enquote {\bibinfo {title} {Buckling
  transitions of an elastic filament in a viscous stagnation point flow},}\
  }\href@noop {} {\bibfield  {journal} {\bibinfo  {journal} {Physics of
  Fluids}\ }\textbf {\bibinfo {volume} {24}},\ \bibinfo {pages} {123601}
  (\bibinfo {year} {2012})}\BibitemShut {NoStop}%
\bibitem [{\citenamefont {Liu}\ \emph {et~al.}(2018)\citenamefont {Liu},
  \citenamefont {Chakrabarti}, \citenamefont {Saintillan}, \citenamefont
  {Lindner},\ and\ \citenamefont {Du~Roure}}]{liu2018morphological}%
  \BibitemOpen
  \bibfield  {author} {\bibinfo {author} {\bibfnamefont {Yanan}\ \bibnamefont
  {Liu}}, \bibinfo {author} {\bibfnamefont {Brato}\ \bibnamefont
  {Chakrabarti}}, \bibinfo {author} {\bibfnamefont {David}\ \bibnamefont
  {Saintillan}}, \bibinfo {author} {\bibfnamefont {Anke}\ \bibnamefont
  {Lindner}}, \ and\ \bibinfo {author} {\bibfnamefont {Olivia}\ \bibnamefont
  {Du~Roure}},\ }\bibfield  {title} {\enquote {\bibinfo {title} {Morphological
  transitions of elastic filaments in shear flow},}\ }\href@noop {} {\bibfield
  {journal} {\bibinfo  {journal} {Proceedings of the National Academy of
  Sciences}\ }\textbf {\bibinfo {volume} {115}},\ \bibinfo {pages} {9438--9443}
  (\bibinfo {year} {2018})}\BibitemShut {NoStop}%
\bibitem [{\citenamefont {LaGrone}\ \emph {et~al.}(2019)\citenamefont
  {LaGrone}, \citenamefont {Cortez}, \citenamefont {Yan},\ and\ \citenamefont
  {Fauci}}]{lagrone2019complex}%
  \BibitemOpen
  \bibfield  {author} {\bibinfo {author} {\bibfnamefont {John}\ \bibnamefont
  {LaGrone}}, \bibinfo {author} {\bibfnamefont {Ricardo}\ \bibnamefont
  {Cortez}}, \bibinfo {author} {\bibfnamefont {Wen}\ \bibnamefont {Yan}}, \
  and\ \bibinfo {author} {\bibfnamefont {Lisa}\ \bibnamefont {Fauci}},\
  }\bibfield  {title} {\enquote {\bibinfo {title} {Complex dynamics of long,
  flexible fibers in shear},}\ }\href@noop {} {\bibfield  {journal} {\bibinfo
  {journal} {Journal of Non-Newtonian Fluid Mechanics}\ }\textbf {\bibinfo
  {volume} {269}},\ \bibinfo {pages} {73--81} (\bibinfo {year}
  {2019})}\BibitemShut {NoStop}%
\bibitem [{\citenamefont {Slowicka}\ \emph {et~al.}(2019)\citenamefont
  {Slowicka}, \citenamefont {Stone},\ and\ \citenamefont
  {Ekiel-Jezewska}}]{slowicka2019flexible}%
  \BibitemOpen
  \bibfield  {author} {\bibinfo {author} {\bibfnamefont {AM}~\bibnamefont
  {Slowicka}}, \bibinfo {author} {\bibfnamefont {Howard~A}\ \bibnamefont
  {Stone}}, \ and\ \bibinfo {author} {\bibfnamefont {Maria~L}\ \bibnamefont
  {Ekiel-Jezewska}},\ }\bibfield  {title} {\enquote {\bibinfo {title} {Flexible
  fibers in shear flow: attracting periodic solutions},}\ }\href@noop {}
  {\bibfield  {journal} {\bibinfo  {journal} {arXiv preprint arXiv:1905.12985}\
  } (\bibinfo {year} {2019})}\BibitemShut {NoStop}%
\bibitem [{\citenamefont {{\.Z}uk}\ \emph {et~al.}(2021)\citenamefont
  {{\.Z}uk}, \citenamefont {S{\l}owicka}, \citenamefont {Ekiel-Je{\.z}ewska},\
  and\ \citenamefont {Stone}}]{zuk2021universal}%
  \BibitemOpen
  \bibfield  {author} {\bibinfo {author} {\bibfnamefont {Pawe{\l}~J}\
  \bibnamefont {{\.Z}uk}}, \bibinfo {author} {\bibfnamefont {Agnieszka~M}\
  \bibnamefont {S{\l}owicka}}, \bibinfo {author} {\bibfnamefont {Maria~L}\
  \bibnamefont {Ekiel-Je{\.z}ewska}}, \ and\ \bibinfo {author} {\bibfnamefont
  {Howard~A}\ \bibnamefont {Stone}},\ }\bibfield  {title} {\enquote {\bibinfo
  {title} {Universal features of the shape of elastic fibres in shear flow},}\
  }\href@noop {} {\bibfield  {journal} {\bibinfo  {journal} {Journal of Fluid
  Mechanics}\ }\textbf {\bibinfo {volume} {914}} (\bibinfo {year}
  {2021})}\BibitemShut {NoStop}%
\bibitem [{\citenamefont {Kuei}\ \emph {et~al.}(2015)\citenamefont {Kuei},
  \citenamefont {S{\l}owicka}, \citenamefont {Ekiel-Je{\.z}ewska},
  \citenamefont {Wajnryb},\ and\ \citenamefont {Stone}}]{kuei2015dynamics}%
  \BibitemOpen
  \bibfield  {author} {\bibinfo {author} {\bibfnamefont {Steve}\ \bibnamefont
  {Kuei}}, \bibinfo {author} {\bibfnamefont {Agnieszka~M}\ \bibnamefont
  {S{\l}owicka}}, \bibinfo {author} {\bibfnamefont {Maria~L}\ \bibnamefont
  {Ekiel-Je{\.z}ewska}}, \bibinfo {author} {\bibfnamefont {Eligiusz}\
  \bibnamefont {Wajnryb}}, \ and\ \bibinfo {author} {\bibfnamefont {Howard~A}\
  \bibnamefont {Stone}},\ }\bibfield  {title} {\enquote {\bibinfo {title}
  {Dynamics and topology of a flexible chain: knots in steady shear flow},}\
  }\href@noop {} {\bibfield  {journal} {\bibinfo  {journal} {New Journal of
  Physics}\ }\textbf {\bibinfo {volume} {17}},\ \bibinfo {pages} {053009}
  (\bibinfo {year} {2015})}\BibitemShut {NoStop}%
\bibitem [{\citenamefont {Hu}\ \emph {et~al.}(2021)\citenamefont {Hu},
  \citenamefont {Chu}, \citenamefont {Shelley},\ and\ \citenamefont
  {Zhang}}]{hu2021levy}%
  \BibitemOpen
  \bibfield  {author} {\bibinfo {author} {\bibfnamefont {Shi-Yuan}\
  \bibnamefont {Hu}}, \bibinfo {author} {\bibfnamefont {Jun-Jun}\ \bibnamefont
  {Chu}}, \bibinfo {author} {\bibfnamefont {Michael~J}\ \bibnamefont
  {Shelley}}, \ and\ \bibinfo {author} {\bibfnamefont {Jun}\ \bibnamefont
  {Zhang}},\ }\bibfield  {title} {\enquote {\bibinfo {title} {L{\'e}vy walks
  and path chaos in the dispersal of elongated structures moving across
  cellular vortical flows},}\ }\href@noop {} {\bibfield  {journal} {\bibinfo
  {journal} {Physical Review Letters}\ }\textbf {\bibinfo {volume} {127}},\
  \bibinfo {pages} {074503} (\bibinfo {year} {2021})}\BibitemShut {NoStop}%
\bibitem [{\citenamefont {Chakrabarti}\ \emph {et~al.}(2020)\citenamefont
  {Chakrabarti}, \citenamefont {Liu}, \citenamefont {LaGrone}, \citenamefont
  {Cortez}, \citenamefont {Fauci}, \citenamefont {du~Roure}, \citenamefont
  {Saintillan},\ and\ \citenamefont {Lindner}}]{chakrabarti2020flexible}%
  \BibitemOpen
  \bibfield  {author} {\bibinfo {author} {\bibfnamefont {Brato}\ \bibnamefont
  {Chakrabarti}}, \bibinfo {author} {\bibfnamefont {Yanan}\ \bibnamefont
  {Liu}}, \bibinfo {author} {\bibfnamefont {John}\ \bibnamefont {LaGrone}},
  \bibinfo {author} {\bibfnamefont {Ricardo}\ \bibnamefont {Cortez}}, \bibinfo
  {author} {\bibfnamefont {Lisa}\ \bibnamefont {Fauci}}, \bibinfo {author}
  {\bibfnamefont {Olivia}\ \bibnamefont {du~Roure}}, \bibinfo {author}
  {\bibfnamefont {David}\ \bibnamefont {Saintillan}}, \ and\ \bibinfo {author}
  {\bibfnamefont {Anke}\ \bibnamefont {Lindner}},\ }\bibfield  {title}
  {\enquote {\bibinfo {title} {Flexible filaments buckle into helicoidal shapes
  in strong compressional flows},}\ }\href@noop {} {\bibfield  {journal}
  {\bibinfo  {journal} {Nature Physics}\ ,\ \bibinfo {pages} {1--6}} (\bibinfo
  {year} {2020})}\BibitemShut {NoStop}%
\bibitem [{\citenamefont {Wiggins}\ \emph {et~al.}(1998)\citenamefont
  {Wiggins}, \citenamefont {Riveline}, \citenamefont {Ott},\ and\ \citenamefont
  {Goldstein}}]{wiggins1998trapping}%
  \BibitemOpen
  \bibfield  {author} {\bibinfo {author} {\bibfnamefont {Chris~H}\ \bibnamefont
  {Wiggins}}, \bibinfo {author} {\bibfnamefont {D}~\bibnamefont {Riveline}},
  \bibinfo {author} {\bibfnamefont {Albrecht}\ \bibnamefont {Ott}}, \ and\
  \bibinfo {author} {\bibfnamefont {Raymond~E}\ \bibnamefont {Goldstein}},\
  }\bibfield  {title} {\enquote {\bibinfo {title} {Trapping and wiggling:
  elastohydrodynamics of driven microfilaments},}\ }\href@noop {} {\bibfield
  {journal} {\bibinfo  {journal} {Biophysical journal}\ }\textbf {\bibinfo
  {volume} {74}},\ \bibinfo {pages} {1043--1060} (\bibinfo {year}
  {1998})}\BibitemShut {NoStop}%
\bibitem [{\citenamefont {Wolgemuth}\ \emph {et~al.}(2000)\citenamefont
  {Wolgemuth}, \citenamefont {Powers},\ and\ \citenamefont
  {Goldstein}}]{wolgemuth2000twirling}%
  \BibitemOpen
  \bibfield  {author} {\bibinfo {author} {\bibfnamefont {Charles~W}\
  \bibnamefont {Wolgemuth}}, \bibinfo {author} {\bibfnamefont {Thomas~R}\
  \bibnamefont {Powers}}, \ and\ \bibinfo {author} {\bibfnamefont {Raymond~E}\
  \bibnamefont {Goldstein}},\ }\bibfield  {title} {\enquote {\bibinfo {title}
  {Twirling and whirling: Viscous dynamics of rotating elastic filaments},}\
  }\href@noop {} {\bibfield  {journal} {\bibinfo  {journal} {Physical Review
  Letters}\ }\textbf {\bibinfo {volume} {84}},\ \bibinfo {pages} {1623}
  (\bibinfo {year} {2000})}\BibitemShut {NoStop}%
\bibitem [{\citenamefont {Lim}\ and\ \citenamefont
  {Peskin}(2004)}]{lim2004simulations}%
  \BibitemOpen
  \bibfield  {author} {\bibinfo {author} {\bibfnamefont {Sookkyung}\
  \bibnamefont {Lim}}\ and\ \bibinfo {author} {\bibfnamefont {Charles~S}\
  \bibnamefont {Peskin}},\ }\bibfield  {title} {\enquote {\bibinfo {title}
  {Simulations of the whirling instability by the immersed boundary method},}\
  }\href@noop {} {\bibfield  {journal} {\bibinfo  {journal} {SIAM Journal on
  Scientific Computing}\ }\textbf {\bibinfo {volume} {25}},\ \bibinfo {pages}
  {2066--2083} (\bibinfo {year} {2004})}\BibitemShut {NoStop}%
\bibitem [{\citenamefont {Wada}\ and\ \citenamefont
  {Netz}(2006)}]{wada2006non}%
  \BibitemOpen
  \bibfield  {author} {\bibinfo {author} {\bibfnamefont {Hirofumi}\
  \bibnamefont {Wada}}\ and\ \bibinfo {author} {\bibfnamefont {Roland~R}\
  \bibnamefont {Netz}},\ }\bibfield  {title} {\enquote {\bibinfo {title}
  {Non-equilibrium hydrodynamics of a rotating filament},}\ }\href@noop {}
  {\bibfield  {journal} {\bibinfo  {journal} {EPL (Europhysics Letters)}\
  }\textbf {\bibinfo {volume} {75}},\ \bibinfo {pages} {645} (\bibinfo {year}
  {2006})}\BibitemShut {NoStop}%
\bibitem [{\citenamefont {Bonacci}\ \emph {et~al.}(2022)\citenamefont
  {Bonacci}, \citenamefont {Chakrabarti}, \citenamefont {Saintillan},
  \citenamefont {Roure},\ and\ \citenamefont {Lindner}}]{bonacci2022dynamics}%
  \BibitemOpen
  \bibfield  {author} {\bibinfo {author} {\bibfnamefont {Francesco}\
  \bibnamefont {Bonacci}}, \bibinfo {author} {\bibfnamefont {Brato}\
  \bibnamefont {Chakrabarti}}, \bibinfo {author} {\bibfnamefont {David}\
  \bibnamefont {Saintillan}}, \bibinfo {author} {\bibfnamefont {Olivia~du}\
  \bibnamefont {Roure}}, \ and\ \bibinfo {author} {\bibfnamefont {Anke}\
  \bibnamefont {Lindner}},\ }\bibfield  {title} {\enquote {\bibinfo {title}
  {Dynamics of flexible filaments in oscillatory shear flows},}\ }\href@noop {}
  {\bibfield  {journal} {\bibinfo  {journal} {arXiv preprint arXiv:2205.08361}\
  } (\bibinfo {year} {2022})}\BibitemShut {NoStop}%
\bibitem [{\citenamefont {Agrawal}\ and\ \citenamefont
  {Mitra}(2022)}]{agrawal2022chaos}%
  \BibitemOpen
  \bibfield  {author} {\bibinfo {author} {\bibfnamefont {Vipin}\ \bibnamefont
  {Agrawal}}\ and\ \bibinfo {author} {\bibfnamefont {Dhrubaditya}\ \bibnamefont
  {Mitra}},\ }\bibfield  {title} {\enquote {\bibinfo {title} {Chaos and
  irreversibility of a flexible filament in periodically driven stokes flow},}\
  }\href@noop {} {\bibfield  {journal} {\bibinfo  {journal} {Physical Review
  E}\ }\textbf {\bibinfo {volume} {106}},\ \bibinfo {pages} {025103} (\bibinfo
  {year} {2022})}\BibitemShut {NoStop}%
\bibitem [{\citenamefont {Krishnamurthy}\ and\ \citenamefont
  {Prakash}(2022)}]{krishnamurthy2022emergent}%
  \BibitemOpen
  \bibfield  {author} {\bibinfo {author} {\bibfnamefont {Deepak}\ \bibnamefont
  {Krishnamurthy}}\ and\ \bibinfo {author} {\bibfnamefont {Manu}\ \bibnamefont
  {Prakash}},\ }\bibfield  {title} {\enquote {\bibinfo {title} {Emergent
  programmable behavior and chaos in dynamically driven active filaments},}\
  }\href@noop {} {\bibfield  {journal} {\bibinfo  {journal} {bioRxiv}\ }
  (\bibinfo {year} {2022})}\BibitemShut {NoStop}%
\bibitem [{\citenamefont {Doi}\ and\ \citenamefont
  {Edwards}(1986)}]{Doi1986theory}%
  \BibitemOpen
  \bibfield  {author} {\bibinfo {author} {\bibfnamefont {M}~\bibnamefont
  {Doi}}\ and\ \bibinfo {author} {\bibfnamefont {SF}~\bibnamefont {Edwards}},\
  }\bibfield  {title} {\enquote {\bibinfo {title} {The theory of polymer
  dynamics},}\ }\href@noop {} {\  (\bibinfo {year} {1986})}\BibitemShut
  {NoStop}%
\bibitem [{\citenamefont {Goldstein}\ and\ \citenamefont
  {Langer}(1995)}]{goldstein1995nonlinear}%
  \BibitemOpen
  \bibfield  {author} {\bibinfo {author} {\bibfnamefont {Raymond~E}\
  \bibnamefont {Goldstein}}\ and\ \bibinfo {author} {\bibfnamefont {Stephen~A}\
  \bibnamefont {Langer}},\ }\bibfield  {title} {\enquote {\bibinfo {title}
  {Nonlinear dynamics of stiff polymers},}\ }\href@noop {} {\bibfield
  {journal} {\bibinfo  {journal} {Physical review letters}\ }\textbf {\bibinfo
  {volume} {75}},\ \bibinfo {pages} {1094} (\bibinfo {year}
  {1995})}\BibitemShut {NoStop}%
\bibitem [{\citenamefont {Goldstein}\ \emph {et~al.}(1998)\citenamefont
  {Goldstein}, \citenamefont {Powers},\ and\ \citenamefont
  {Wiggins}}]{goldstein1998viscous}%
  \BibitemOpen
  \bibfield  {author} {\bibinfo {author} {\bibfnamefont {Raymond~E}\
  \bibnamefont {Goldstein}}, \bibinfo {author} {\bibfnamefont {Thomas~R}\
  \bibnamefont {Powers}}, \ and\ \bibinfo {author} {\bibfnamefont {Chris~H}\
  \bibnamefont {Wiggins}},\ }\bibfield  {title} {\enquote {\bibinfo {title}
  {Viscous nonlinear dynamics of twist and writhe},}\ }\href@noop {} {\bibfield
   {journal} {\bibinfo  {journal} {Physical Review Letters}\ }\textbf {\bibinfo
  {volume} {80}},\ \bibinfo {pages} {5232} (\bibinfo {year}
  {1998})}\BibitemShut {NoStop}%
\bibitem [{\citenamefont {Alligood}\ \emph {et~al.}(1996)\citenamefont
  {Alligood}, \citenamefont {Sauer},\ and\ \citenamefont
  {Yorke}}]{Alligood1996chaos}%
  \BibitemOpen
  \bibfield  {author} {\bibinfo {author} {\bibfnamefont {Kathleen~T}\
  \bibnamefont {Alligood}}, \bibinfo {author} {\bibfnamefont {Tim~D}\
  \bibnamefont {Sauer}}, \ and\ \bibinfo {author} {\bibfnamefont {James~A}\
  \bibnamefont {Yorke}},\ }\href@noop {} {\emph {\bibinfo {title} {Chaos: An
  introduction to dynamical systems}}}\ (\bibinfo  {publisher} {Springer},\
  \bibinfo {address} {New York},\ \bibinfo {year} {1996})\BibitemShut {NoStop}%
\bibitem [{\citenamefont {Larson}\ \emph {et~al.}(1999)\citenamefont {Larson},
  \citenamefont {Hu}, \citenamefont {Smith},\ and\ \citenamefont
  {Chu}}]{larson1999brownian}%
  \BibitemOpen
  \bibfield  {author} {\bibinfo {author} {\bibfnamefont {RG}~\bibnamefont
  {Larson}}, \bibinfo {author} {\bibfnamefont {Hua}\ \bibnamefont {Hu}},
  \bibinfo {author} {\bibfnamefont {DE}~\bibnamefont {Smith}}, \ and\ \bibinfo
  {author} {\bibfnamefont {S}~\bibnamefont {Chu}},\ }\bibfield  {title}
  {\enquote {\bibinfo {title} {Brownian dynamics simulations of a dna molecule
  in an extensional flow field},}\ }\href@noop {} {\bibfield  {journal}
  {\bibinfo  {journal} {Journal of Rheology}\ }\textbf {\bibinfo {volume}
  {43}},\ \bibinfo {pages} {267--304} (\bibinfo {year} {1999})}\BibitemShut
  {NoStop}%
\bibitem [{\citenamefont {Nazockdast}\ \emph {et~al.}(2017)\citenamefont
  {Nazockdast}, \citenamefont {Rahimian}, \citenamefont {Zorin},\ and\
  \citenamefont {Shelley}}]{nazockdast2017fast}%
  \BibitemOpen
  \bibfield  {author} {\bibinfo {author} {\bibfnamefont {Ehssan}\ \bibnamefont
  {Nazockdast}}, \bibinfo {author} {\bibfnamefont {Abtin}\ \bibnamefont
  {Rahimian}}, \bibinfo {author} {\bibfnamefont {Denis}\ \bibnamefont {Zorin}},
  \ and\ \bibinfo {author} {\bibfnamefont {Michael}\ \bibnamefont {Shelley}},\
  }\bibfield  {title} {\enquote {\bibinfo {title} {A fast platform for
  simulating semi-flexible fiber suspensions applied to cell mechanics},}\
  }\href@noop {} {\bibfield  {journal} {\bibinfo  {journal} {Journal of
  Computational Physics}\ }\textbf {\bibinfo {volume} {329}},\ \bibinfo {pages}
  {173--209} (\bibinfo {year} {2017})}\BibitemShut {NoStop}%
\bibitem [{\citenamefont {Wada}\ and\ \citenamefont
  {Netz}(2007)}]{wada2007stretching}%
  \BibitemOpen
  \bibfield  {author} {\bibinfo {author} {\bibfnamefont {Hirofumi}\
  \bibnamefont {Wada}}\ and\ \bibinfo {author} {\bibfnamefont {Roland~R}\
  \bibnamefont {Netz}},\ }\bibfield  {title} {\enquote {\bibinfo {title}
  {Stretching helical nano-springs at finite temperature},}\ }\href@noop {}
  {\bibfield  {journal} {\bibinfo  {journal} {EPL (Europhysics Letters)}\
  }\textbf {\bibinfo {volume} {77}},\ \bibinfo {pages} {68001} (\bibinfo {year}
  {2007})}\BibitemShut {NoStop}%
\bibitem [{\citenamefont {Powers}(2010)}]{powers2010dynamics}%
  \BibitemOpen
  \bibfield  {author} {\bibinfo {author} {\bibfnamefont {Thomas~R}\
  \bibnamefont {Powers}},\ }\bibfield  {title} {\enquote {\bibinfo {title}
  {Dynamics of filaments and membranes in a viscous fluid},}\ }\href@noop {}
  {\bibfield  {journal} {\bibinfo  {journal} {Reviews of Modern Physics}\
  }\textbf {\bibinfo {volume} {82}},\ \bibinfo {pages} {1607} (\bibinfo {year}
  {2010})}\BibitemShut {NoStop}%
\bibitem [{\citenamefont {Press}\ and\ \citenamefont
  {Teukolsky}(1992)}]{press1992adaptive}%
  \BibitemOpen
  \bibfield  {author} {\bibinfo {author} {\bibfnamefont {William~H}\
  \bibnamefont {Press}}\ and\ \bibinfo {author} {\bibfnamefont {Saul~A}\
  \bibnamefont {Teukolsky}},\ }\bibfield  {title} {\enquote {\bibinfo {title}
  {Adaptive stepsize runge-kutta integration},}\ }\href@noop {} {\bibfield
  {journal} {\bibinfo  {journal} {Computers in Physics}\ }\textbf {\bibinfo
  {volume} {6}},\ \bibinfo {pages} {188--191} (\bibinfo {year}
  {1992})}\BibitemShut {NoStop}%
\bibitem [{\citenamefont {Cash}\ and\ \citenamefont
  {Karp}(1990)}]{cash1990variable}%
  \BibitemOpen
  \bibfield  {author} {\bibinfo {author} {\bibfnamefont {Jeff~R}\ \bibnamefont
  {Cash}}\ and\ \bibinfo {author} {\bibfnamefont {Alan~H}\ \bibnamefont
  {Karp}},\ }\bibfield  {title} {\enquote {\bibinfo {title} {A variable order
  runge-kutta method for initial value problems with rapidly varying right-hand
  sides},}\ }\href@noop {} {\bibfield  {journal} {\bibinfo  {journal} {ACM
  Transactions on Mathematical Software (TOMS)}\ }\textbf {\bibinfo {volume}
  {16}},\ \bibinfo {pages} {201--222} (\bibinfo {year} {1990})}\BibitemShut
  {NoStop}%
\bibitem [{Note1()}]{Note1}%
  \BibitemOpen
  \bibinfo {note} {\protect \url
  {https://github.com/dhrubaditya/ElasticString}}\BibitemShut {NoStop}%
\bibitem [{\citenamefont {Ramamohan}\ \emph {et~al.}(1994)\citenamefont
  {Ramamohan}, \citenamefont {Savithri}, \citenamefont {Sreenivasan},\ and\
  \citenamefont {Bhat}}]{ramamohan1994chaotic}%
  \BibitemOpen
  \bibfield  {author} {\bibinfo {author} {\bibfnamefont {TR}~\bibnamefont
  {Ramamohan}}, \bibinfo {author} {\bibfnamefont {S}~\bibnamefont {Savithri}},
  \bibinfo {author} {\bibfnamefont {R}~\bibnamefont {Sreenivasan}}, \ and\
  \bibinfo {author} {\bibfnamefont {C~Chandra~Shekara}\ \bibnamefont {Bhat}},\
  }\bibfield  {title} {\enquote {\bibinfo {title} {Chaotic dynamics of a
  periodically forced slender body in a simple shear flow},}\ }\href@noop {}
  {\bibfield  {journal} {\bibinfo  {journal} {Physics Letters A}\ }\textbf
  {\bibinfo {volume} {190}},\ \bibinfo {pages} {273--278} (\bibinfo {year}
  {1994})}\BibitemShut {NoStop}%
\bibitem [{\citenamefont {Kumar}\ \emph {et~al.}(1995)\citenamefont {Kumar},
  \citenamefont {Kumar},\ and\ \citenamefont {Ramamohan}}]{kumar1995chaotic}%
  \BibitemOpen
  \bibfield  {author} {\bibinfo {author} {\bibfnamefont {CV}~\bibnamefont
  {Kumar}}, \bibinfo {author} {\bibfnamefont {K~Satheesh}\ \bibnamefont
  {Kumar}}, \ and\ \bibinfo {author} {\bibfnamefont {TR}~\bibnamefont
  {Ramamohan}},\ }\bibfield  {title} {\enquote {\bibinfo {title} {Chaotic
  dynamics of periodically forced spheroids in simple shear flow with potential
  application to particle separation},}\ }\href@noop {} {\bibfield  {journal}
  {\bibinfo  {journal} {Rheologica acta}\ }\textbf {\bibinfo {volume} {34}},\
  \bibinfo {pages} {504--511} (\bibinfo {year} {1995})}\BibitemShut {NoStop}%
\bibitem [{\citenamefont {Lundell}(2011)}]{lundell2011effect}%
  \BibitemOpen
  \bibfield  {author} {\bibinfo {author} {\bibfnamefont {Fredrik}\ \bibnamefont
  {Lundell}},\ }\bibfield  {title} {\enquote {\bibinfo {title} {The effect of
  particle inertia on triaxial ellipsoids in creeping shear: from drift toward
  chaos to a single periodic solution},}\ }\href@noop {} {\bibfield  {journal}
  {\bibinfo  {journal} {Physics of Fluids}\ }\textbf {\bibinfo {volume} {23}},\
  \bibinfo {pages} {011704} (\bibinfo {year} {2011})}\BibitemShut {NoStop}%
\bibitem [{\citenamefont {Nilsen}\ and\ \citenamefont
  {Andersson}(2013)}]{nilsen2013chaotic}%
  \BibitemOpen
  \bibfield  {author} {\bibinfo {author} {\bibfnamefont {Christopher}\
  \bibnamefont {Nilsen}}\ and\ \bibinfo {author} {\bibfnamefont {Helge~I}\
  \bibnamefont {Andersson}},\ }\bibfield  {title} {\enquote {\bibinfo {title}
  {Chaotic rotation of inertial spheroids in oscillating shear flow},}\
  }\href@noop {} {\bibfield  {journal} {\bibinfo  {journal} {Physics of
  Fluids}\ }\textbf {\bibinfo {volume} {25}},\ \bibinfo {pages} {013303}
  (\bibinfo {year} {2013})}\BibitemShut {NoStop}%
\bibitem [{\citenamefont {Auerbach}\ \emph {et~al.}(1987)\citenamefont
  {Auerbach}, \citenamefont {Cvitanovi{\'c}}, \citenamefont {Eckmann},
  \citenamefont {Gunaratne},\ and\ \citenamefont
  {Procaccia}}]{auerbach1987exploring}%
  \BibitemOpen
  \bibfield  {author} {\bibinfo {author} {\bibfnamefont {Ditza}\ \bibnamefont
  {Auerbach}}, \bibinfo {author} {\bibfnamefont {Predrag}\ \bibnamefont
  {Cvitanovi{\'c}}}, \bibinfo {author} {\bibfnamefont {Jean-Pierre}\
  \bibnamefont {Eckmann}}, \bibinfo {author} {\bibfnamefont {Gemunu}\
  \bibnamefont {Gunaratne}}, \ and\ \bibinfo {author} {\bibfnamefont {Itamar}\
  \bibnamefont {Procaccia}},\ }\bibfield  {title} {\enquote {\bibinfo {title}
  {Exploring chaotic motion through periodic orbits},}\ }\href@noop {}
  {\bibfield  {journal} {\bibinfo  {journal} {Physical Review Letters}\
  }\textbf {\bibinfo {volume} {58}},\ \bibinfo {pages} {2387} (\bibinfo {year}
  {1987})}\BibitemShut {NoStop}%
\bibitem [{\citenamefont {Cvitanovic}\ \emph {et~al.}(2005)\citenamefont
  {Cvitanovic}, \citenamefont {Artuso}, \citenamefont {Mainieri}, \citenamefont
  {Tanner}, \citenamefont {Vattay}, \citenamefont {Whelan},\ and\ \citenamefont
  {Wirzba}}]{cvitanovic2005chaos}%
  \BibitemOpen
  \bibfield  {author} {\bibinfo {author} {\bibfnamefont {Predrag}\ \bibnamefont
  {Cvitanovic}}, \bibinfo {author} {\bibfnamefont {Roberto}\ \bibnamefont
  {Artuso}}, \bibinfo {author} {\bibfnamefont {Ronnie}\ \bibnamefont
  {Mainieri}}, \bibinfo {author} {\bibfnamefont {Gregor}\ \bibnamefont
  {Tanner}}, \bibinfo {author} {\bibfnamefont {G{\'a}bor}\ \bibnamefont
  {Vattay}}, \bibinfo {author} {\bibfnamefont {Niall}\ \bibnamefont {Whelan}},
  \ and\ \bibinfo {author} {\bibfnamefont {Andreas}\ \bibnamefont {Wirzba}},\
  }\bibfield  {title} {\enquote {\bibinfo {title} {Chaos: classical and
  quantum},}\ }\href@noop {} {\bibfield  {journal} {\bibinfo  {journal}
  {ChaosBook. org (Niels Bohr Institute, Copenhagen 2005)}\ }\textbf {\bibinfo
  {volume} {69}},\ \bibinfo {pages} {25} (\bibinfo {year} {2005})}\BibitemShut
  {NoStop}%
\bibitem [{\citenamefont {Sharkovski}(1995)}]{sharkovskiui1995coexistence}%
  \BibitemOpen
  \bibfield  {author} {\bibinfo {author} {\bibfnamefont {AN}~\bibnamefont
  {Sharkovski}},\ }\bibfield  {title} {\enquote {\bibinfo {title} {Coexistence
  of cycles of a continuous map of the line into itself},}\ }\href@noop {}
  {\bibfield  {journal} {\bibinfo  {journal} {International journal of
  bifurcation and chaos}\ }\textbf {\bibinfo {volume} {5}},\ \bibinfo {pages}
  {1263--1273} (\bibinfo {year} {1995})}\BibitemShut {NoStop}%
\bibitem [{\citenamefont {Li}\ and\ \citenamefont
  {Yorke}(2004)}]{li2004period}%
  \BibitemOpen
  \bibfield  {author} {\bibinfo {author} {\bibfnamefont {Tien-Yien}\
  \bibnamefont {Li}}\ and\ \bibinfo {author} {\bibfnamefont {James~A}\
  \bibnamefont {Yorke}},\ }\bibfield  {title} {\enquote {\bibinfo {title}
  {Period three implies chaos},}\ }in\ \href@noop {} {\emph {\bibinfo
  {booktitle} {The theory of chaotic attractors}}}\ (\bibinfo  {publisher}
  {Springer},\ \bibinfo {year} {2004})\ pp.\ \bibinfo {pages}
  {77--84}\BibitemShut {NoStop}%
\bibitem [{Note2()}]{Note2}%
  \BibitemOpen
  \bibinfo {note} {As a counterexample~\cite {kloeden2006li}, consider the two
  dimensional map that rotates every point in the $x-y$ plane by an angle of
  $2\pi /3$ in the counter-clockwise direction. Clearly this map has a period
  three solution but it is not chaotic.}\BibitemShut {Stop}%
\bibitem [{Note3()}]{Note3}%
  \BibitemOpen
  \bibinfo {note} {It is also possible that the shape shows period three
  solution but the $\Theta $ shows a fixed point or a period--two solution.
  Conversely, it is possible for $\Theta $ to have a period--three solution
  without the shape having a period-three solution.}\BibitemShut {Stop}%
\bibitem [{\citenamefont {Rotne}\ and\ \citenamefont
  {Prager}(1969)}]{rotne1969variational}%
  \BibitemOpen
  \bibfield  {author} {\bibinfo {author} {\bibfnamefont {Jens}\ \bibnamefont
  {Rotne}}\ and\ \bibinfo {author} {\bibfnamefont {Stephen}\ \bibnamefont
  {Prager}},\ }\bibfield  {title} {\enquote {\bibinfo {title} {Variational
  treatment of hydrodynamic interaction in polymers},}\ }\href@noop {}
  {\bibfield  {journal} {\bibinfo  {journal} {The Journal of Chemical Physics}\
  }\textbf {\bibinfo {volume} {50}},\ \bibinfo {pages} {4831--4837} (\bibinfo
  {year} {1969})}\BibitemShut {NoStop}%
\bibitem [{\citenamefont {Brady}\ and\ \citenamefont
  {Bossis}(1988)}]{brady1988stokesian}%
  \BibitemOpen
  \bibfield  {author} {\bibinfo {author} {\bibfnamefont {John~F}\ \bibnamefont
  {Brady}}\ and\ \bibinfo {author} {\bibfnamefont {Georges}\ \bibnamefont
  {Bossis}},\ }\bibfield  {title} {\enquote {\bibinfo {title} {Stokesian
  dynamics},}\ }\href@noop {} {\bibfield  {journal} {\bibinfo  {journal}
  {Annual review of fluid mechanics}\ }\textbf {\bibinfo {volume} {20}},\
  \bibinfo {pages} {111--157} (\bibinfo {year} {1988})}\BibitemShut {NoStop}%
\bibitem [{\citenamefont {Guazzelli}\ and\ \citenamefont
  {Morris}(2011)}]{guazzelli2011physical}%
  \BibitemOpen
  \bibfield  {author} {\bibinfo {author} {\bibfnamefont {Elisabeth}\
  \bibnamefont {Guazzelli}}\ and\ \bibinfo {author} {\bibfnamefont {Jeffrey~F}\
  \bibnamefont {Morris}},\ }\href@noop {} {\emph {\bibinfo {title} {A physical
  introduction to suspension dynamics}}},\ Vol.~\bibinfo {volume} {45}\
  (\bibinfo  {publisher} {Cambridge University Press},\ \bibinfo {year}
  {2011})\BibitemShut {NoStop}%
\bibitem [{\citenamefont {Kim}\ and\ \citenamefont
  {Karrila}(2013)}]{kim2013microhydrodynamics}%
  \BibitemOpen
  \bibfield  {author} {\bibinfo {author} {\bibfnamefont {Sangtae}\ \bibnamefont
  {Kim}}\ and\ \bibinfo {author} {\bibfnamefont {Seppo~J}\ \bibnamefont
  {Karrila}},\ }\href@noop {} {\emph {\bibinfo {title} {Microhydrodynamics:
  principles and selected applications}}}\ (\bibinfo  {publisher} {Courier
  Corporation},\ \bibinfo {year} {2013})\BibitemShut {NoStop}%
\bibitem [{\citenamefont {Taylor}(1922)}]{taylor1922diffusion}%
  \BibitemOpen
  \bibfield  {author} {\bibinfo {author} {\bibfnamefont {Geoffrey~I}\
  \bibnamefont {Taylor}},\ }\bibfield  {title} {\enquote {\bibinfo {title}
  {Diffusion by continuous movements},}\ }\href@noop {} {\bibfield  {journal}
  {\bibinfo  {journal} {Proceedings of the london mathematical society}\
  }\textbf {\bibinfo {volume} {2}},\ \bibinfo {pages} {196--212} (\bibinfo
  {year} {1922})}\BibitemShut {NoStop}%
\bibitem [{\citenamefont {Kloeden}\ and\ \citenamefont
  {Li}(2006)}]{kloeden2006li}%
  \BibitemOpen
  \bibfield  {author} {\bibinfo {author} {\bibfnamefont {Peter}\ \bibnamefont
  {Kloeden}}\ and\ \bibinfo {author} {\bibfnamefont {Zhong}\ \bibnamefont
  {Li}},\ }\bibfield  {title} {\enquote {\bibinfo {title} {Li--yorke chaos in
  higher dimensions: A review},}\ }\href@noop {} {\bibfield  {journal}
  {\bibinfo  {journal} {Journal of Difference Equations and Applications}\
  }\textbf {\bibinfo {volume} {12}},\ \bibinfo {pages} {247--269} (\bibinfo
  {year} {2006})}\BibitemShut {NoStop}%
\end{thebibliography}
